\documentclass[11pt]{article}
\usepackage{amsmath,amsthm,amssymb,cite,enumerate}
\usepackage{multirow}
\usepackage[a4paper,left=0.5in,right=1in]{geometry}
\usepackage[usenames]{color}
\usepackage{graphicx}
\usepackage{epsfig}
\usepackage{geometry}
\usepackage{array} 
\usepackage{mathrsfs}
\usepackage{dcolumn}
\usepackage{bm}
\usepackage{authblk} 
\usepackage{soul} 
\usepackage{t1enc}	
\usepackage{comment}
\usepackage{xcolor}
\usepackage{longtable}
\begin{document}

\def \d {{\rm d}}

\def \bm {\mbox{\boldmath{$m$}}}
\def \bmt {\mbox{\boldmath{$\tilde m$}}}

\def \bF {\mbox{\boldmath{$F$}}}
\def \bA {\mbox{\boldmath{$A$}}}
\def \cF {\mbox{\boldmath{$\cal F$}}}
\def \bH {\mbox{\boldmath{$H$}}}
\def \bC {\mbox{\boldmath{$C$}}}
\def \bSS {\mbox{\boldmath{$S$}}}
\def \bS {\mbox{\boldmath{${\cal S}$}}}
\def \bV {\mbox{\boldmath{$V$}}}
\def \bff {\mbox{\boldmath{$f$}}}
\def \bT {\mbox{\boldmath{$T$}}}
\def \bk {\mbox{\boldmath{$k$}}}
\def \bl {\mbox{\boldmath{$l$}}}
\def \bn {\mbox{\boldmath{$n$}}}
\def \bbm {\mbox{\boldmath{$m$}}}
\def \tbbm {\mbox{\boldmath{$\bar m$}}}
\def \bet {\mbox{\boldmath{$\eta$}}}
\def \H {{\cal H}}
\def \bmu {\mbox{\boldmath{$\mu$}}}
\def \bSig {\mbox{\boldmath{$\Sigma$}}}
\def \T {\bigtriangleup}
\newcommand{\msub}[2]{m^{(#1)}_{#2}}
\newcommand{\msup}[2]{m_{(#1)}^{#2}}

\newcommand{\be}{\begin{equation}}
\newcommand{\ee}{\end{equation}}

\newcommand{\beqn}{\begin{eqnarray}}
\newcommand{\eeqn}{\end{eqnarray}}
\newcommand{\AdS}{anti--de~Sitter }
\newcommand{\AAdS}{\mbox{(anti--)}de~Sitter }
\newcommand{\AAN}{\mbox{(anti--)}Nariai }
\newcommand{\AS}{Aichelburg-Sexl }
\newcommand{\pa}{\partial}
\newcommand{\pp}{{\it pp\,}-}
\newcommand{\ba}{\begin{array}}
\newcommand{\ea}{\end{array}}

\newcommand*\bR{\ensuremath{\boldsymbol{R}}}

\newcommand*\BF{\ensuremath{\boldsymbol{F}}}
\newcommand*\BR{\ensuremath{\boldsymbol{R}}}
\newcommand*\BS{\ensuremath{\boldsymbol{S}}}
\newcommand*\BC{\ensuremath{\boldsymbol{C}}}
\newcommand*\bg{\ensuremath{\boldsymbol{g}}}
\newcommand*\bE{\ensuremath{\boldsymbol{E}}}

\newcommand*\bh{\ensuremath{\boldsymbol{h}}}
\newcommand*\bZ{\ensuremath{\boldsymbol{Z}}}

\def \bo {\mbox{\boldmath{$\omega$}}}
\def \bot {\mbox{\boldmath{$\tilde\omega$}}}
\def \bE {\mbox{\boldmath{$e$}}}
\def \bEt {\mbox{\boldmath{$\tilde e$}}}
\def \bG {\mbox{\boldmath{$\Gamma$}}}
\def \bGt {\mbox{\boldmath{$\tilde\Gamma$}}}
\def \bTt {\mbox{\boldmath{$\tilde\Theta$}}}

\newcommand{\M}[3] {{\stackrel{#1}{M}}_{{#2}{#3}}}
\newcommand{\m}[3] {{\stackrel{\hspace{.3cm}#1}{m}}_{\!{#2}{#3}}\,}

\newcommand{\tr}{\textcolor{red}}
\newcommand{\tb}{\textcolor{blue}}
\newcommand{\tg}{\textcolor{green}}

\newcommand{\thorn}{\mathop{\hbox{\rm \th}}\nolimits}

\def\a{\alpha}
\def\g{\gamma}
\def\de{\delta}

\def\E{{\cal E}}
\def\B{{\cal B}}
\def\R{{\cal R}}
\def\F{{\cal F}}
\def\L{{\cal L}}

\def\e{e}
\def\bb{b}

\newtheorem{theorem}{Theorem}[section] 
\newtheorem{cor}[theorem]{Corollary} 
\newtheorem{lemma}[theorem]{Lemma} 
\newtheorem{proposition}[theorem]{Proposition}
\newtheorem{definition}[theorem]{Definition}
\newtheorem{remark}[theorem]{Remark}  

\title{Charging higher-dimensional spacetimes with a generalized Kerr-Schild transformation}

\author[1,2]{Aravindhan Srinivasan\thanks{srinivasan(at)math(dot)cas(dot)cz}}

\author[2]{Marcello Ortaggio\thanks{ortaggio(at)math(dot)cas(dot)cz}}

\affil[1]{Institute of Theoretical Physics, Faculty of Mathematics and Physics, \newline
 Charles University, V Hole\v{s}ovi\v{c}k\'{a}ch 2, 180 00 Prague 8, Czech Republic}

\affil[2]{Institute of Mathematics, Czech Academy of Sciences, \newline \v Zitn\' a 25, 115 67 Prague 1, Czech Republic}

\maketitle

\abstract{We explore the construction of higher-dimensional Einstein-Maxwell(-Chern-Simons) solutions from vacuum seeds by means of a generalized Kerr-Schild transformation along a geodesic null vector field $\mathbf{k}$. Assuming the vector potential $\mathbf{A}$ to be aligned with $\mathbf{k}$, and $\mathbf{k}$ to be a Weyl aligned null direction satisfying the ``optical constraint'', we arrive at three distinct branches of solutions. If $\mathbf{k}$ is {\em expanding and twisting}, then its shear must vanish -- this branch includes certain charged Taub-NUT metrics. If $\mathbf{k}$ is {\em expanding and twistfree}, one finds two subfamilies:  Robinson-Trautman electrovac solutions with a non-null Maxwell field if the shear is zero, or shearing solutions with a null field. Finally, the case when $\mathbf{k}$ is {\em non-expanding} reduces to a subset of the known Kundt solutions. In all cases, the Chern-Simons term turns out to be identically zero on-shell. In passing, by relaxing the alignment assumption on $\mathbf{A}$, we also obtain a six-dimensional extension of a charged Taub-NUT metric for which the magnetic part of the field strength is a linear combination of two distinct K\"ahler 2-forms, as opposed to the previously known examples in more than four dimensions.}

\vspace{.2cm}
\noindent

\tableofcontents

\section{Introduction}

\label{sec_intro}

\subsection{Background}

\label{subsec_backgr}

The Kerr metric was originally written as a Kerr-Schild (KS) transformation of flat spacetime, namely \cite{Kerr63}
\begin{align}
    \mathbf{g}&=\mathbf{\Bar{g}} -2H \mathbf{k}\otimes \mathbf{k} ,
		\label{GKS}
\end{align}
where $\mathbf{\Bar{g}}$ is the Minkowski metric (in suitable coordinates), the covector $\mathbf{k}$ is null (equivalently w.r.t. $\mathbf{g}$ or $\mathbf{\Bar{g}}$), and $H=H_{\mbox{\tiny Kerr}}$ a specific spacetime function proportional to the mass parameter. Using the same form of $\mathbf{\Bar{g}}$ and $\mathbf{k}$ as in \cite{Kerr63}, the line-element~\eqref{GKS} also describes the Kerr-Newman metric \cite{Newmanetal65}, provided one replaces the function $H=H_{\mbox{\tiny Kerr}}\mapsto H_{\mbox{\tiny KN}}$, where $H_{\mbox{\tiny KN}}$ additionally contains a term proportional to the charge parameter. The corresponding Maxwell field $\mathbf{F}=d\mathbf{A}$ is given by
\begin{align}
	\mathbf{A}=\alpha \mathbf{k} ,
\label{aligned_A}
\end{align}
where $\alpha$ is a function proportional to $H_{\mbox{\tiny Kerr}}$ \cite{Newmanetal65}. Both the Kerr and the Kerr-Newman metric thus belong to the KS class of spacetimes \cite{Trautman62,KerSch652}.

Einstein-Maxwell solutions of the KS class described by functions $H$ and $\alpha$ more general than in the Kerr-Newman case were constructed in \cite{DebKerSch69}.\footnote{It should be observed that the vector potential of the electrovac KS solutions of \cite{DebKerSch69} (see also \cite{Stephanibook} for a review) is not restricted to the form~\eqref{aligned_A}. The latter is recovered (up to gauge) precisely when the function $\psi$ of \cite{DebKerSch69} is taken to be real. See~\cite{Ayon-Beatoetal1025} for a related analysis.} 
 However, a further generalization consists in still adopting the ansatz \eqref{GKS}, \eqref{aligned_A}, but now taking $\mathbf{\Bar{g}}$ to be a non-flat background. In particular, when $\mathbf{\Bar{g}}$ is the (A)dS metric, such an extended ansatz notably encompasses the Kerr-Newman-(A)dS solution \cite{Carter68cmp,Carter73} (see, e.g., our recent discussion in Appendix~A of \cite{Ort_Srini} for more details and references). 
Any two geometries $\mathbf{g}$ and $\mathbf{\Bar{g}}$ satisfying~\eqref{GKS} are said to be related by a {\em generalized Kerr–Schild (GKS) transformation}.\footnote{
The above mentioned Kerr-Newman metric can thus be seen also as a GKS transformation of the Kerr metric (w.r.t. the same $\mathbf{k}$) with a function $H=H_{\mbox{\tiny KN}}-H_{\mbox{\tiny Kerr}}$ proportional to the charge parameter. This is the viewpoint more relevant to the present paper.} 
 Various properties of such transformation (in particular, the relation between the curvature tensors and the Newman-Penrose coefficients of $\mathbf{g}$ and $\mathbf{\Bar{g}}$) were analyzed in detail in \cite{Thompson66,Edelen66_I,DebKerSch69,Dozmorov70,GurGur75,Xanthopoulos78,Taub81,bilge,Xanthopoulos83,MarSen86,BilGur86,Nahmad88} and reviewed in \cite{Stephanibook} (also alerting about some misprints in \cite{bilge}). Group theoretic aspects (in arbitrary spacetime dimensions) were explored in \cite{ColHilSen01}. Numerous references using the GKS transformation as a tool to construct new exact solutions from old ones can be found in \cite{Stephanibook}.

Remarkably, a higher-dimensional extension of the Kerr black hole was obtained by Myers and Perry \cite{MyePer86} using an ansatz of the form~\eqref{GKS}, 
with $\mathbf{\Bar{g}}$ being flat (see also \cite{Chakrabarti86} in eight dimensions). Adding a cosmological constant $\Lambda$ leads to the Einstein spacetimes of\cite{HawHunTay99,Gibbonsetal05}, which can also be cast in the form~\eqref{GKS} \cite{Gibbonsetal05}, with $\mathbf{\Bar{g}}$ now being (A)dS.\footnote{This remains true also for the analytic continuations and limits of the metric of \cite{Gibbonsetal05} considered in \cite{MarPeo22_b,ChrConGra25} (see also \cite{KleMorVan98,Klemm98,deFGodRea15,Ortaggio17,MarPaeSen17,KokOrt25}).} Even prior to the Myers-Perry paper \cite{MyePer86}, the (G)KS tranformation in higher dimensions was studied in \cite{DerGur86}, while more recent analysis includes \cite{ColHilSen01,OrtPraPra09,Malek:2010mh, Ort_Srini, Srinivasan:2025hro}. A simple but remarkable fact is that, if $\mathbf{k}$ is assumed to be {\em geodesic}, the Maxwell equations for the ansatz~\eqref{aligned_A} take the same form in both geometries $\mathbf{g}$ and $\mathbf{\Bar{g}}$, i.e., they do not involve the function $H$ \cite{MyePer86,Ort_Srini}.\footnote{Under~\eqref{aligned_A}, $\mathbf{k}$ being geodesic is equivalent to $\mathbf{F}$ being {\em aligned} with $\mathbf{k}$ (i.e., $F^a_{\phantom{a}b}k^b\propto k^a$ \cite{BerSen01,Coleyetal04vsi,Milson05} or, equivalently, $k^aF_{a[b}k_{c]}=0$ \cite{HerOrtWyl13}). More generally, the Maxwell equations take the same form in both geometries $\mathbf{g}$ and $\mathbf{\Bar{g}}$ for any $\mathbf{F}$ aligned with $\mathbf{k}$ \cite{Harte17}, i.e., even when~\eqref{aligned_A} is not assumed and $\mathbf{k}$ is not geodesic. The result of \cite{Harte17} (which also follows from the expressions given in \cite{BilGur86}, cf. a comment in \cite{Kupeli88}) was proven in four dimensions, but extends readily to any $n$ by noticing that, thanks to~\eqref{GKS} and $k^ak_a=0$, one has $F^{ab}=\Bar{g}^{ac}\Bar{g}^{bd}F_{cd}+4Hk^{[a}\Bar{g}^{b]d}k^cF_{cd}$.} Naturally, this was considered as a first step in the search for an extension of the Kerr-Newman solution to higher dimensions in \cite{MyePer86}. However, the same reference showed that metrics with precisely one non-zero spin do not admit a charged extension within the KS ansatz~\eqref{GKS}, \eqref{aligned_A} with a flat $\mathbf{\Bar{g}}$. Recently, we have extended the no-go result of \cite{MyePer86} to allow $\mathbf{\Bar{g}}$ to be a spacetime of any constant curvature (thus admitting an arbitrary cosmological constant), and $\mathbf{k}$ to define any geodesic, expanding and twisting null congruence \cite{Ort_Srini}. As it turns out, the field equations require $\mathbf{k}$ to be shearfree. This means that one cannot add charge to the rotating black holes of \cite{MyePer86,HawHunTay99,Gibbonsetal05} using a KS transformation~\eqref{GKS}, \eqref{aligned_A}\footnote{Certain charged rotating AdS solutions with a flat horizon have been constructed in \cite{Awad03} (as an extension of the $n=4$ ones of \cite{Lemos95,LemZan96}) but do not contradict our no-go result since they are locally isometric to some of the static solutions given in \cite{GibWil87,KodIsh04,OrtPodZof08}, and possess two twistfree multiple Weyl aligned null directions (mWANDs)). Beyond the KS class, five-dimensional charged Einstein-Maxwell black holes with non-zero angular momentum (and which are not locally static) have been proven to exist \cite{KunNavPer05,KunNavVie06,Frobetal22}. More numerous black holes/rings are known if one includes also an electromagnetic Chern-Simons term in the theory, e.g., \cite{Breckenridgeetal97,CCLP,Elvangetal04,DesLun25} (see Remark~\ref{rem_CCLP} and \cite{Srinivasan:2025hro} for related comments, and \cite{EmpRea08} for more references).} and that the only electrovac solutions of this class are given by a family of higher dimensional charged Taub-NUT metrics with a base space of constant holomorphic sectional curvature (the latter conclusion follows thanks to a result of \cite{Taghavi-Chabert22} -- see also \cite{Ortaggio17} in six dimensions with $\Lambda=0$). These are a subset of solutions previously constructed in \cite{ManSte06,Awad06,DehKoh06}. The case when $\mathbf{k}$ is still expanding but twistfree was also analyzed in \cite{Ort_Srini}.

The main purpose of the present contribution is to extend the results of \cite{Ort_Srini} to the more general case of a (charging) GKS transformation.\footnote{The analysis of \cite{Ort_Srini} encompassed spacetimes of the form~\eqref{GKS} for which $\mathbf{\Bar{g}}$ is either flat or (A)dS (both of which were, for the sake of brevity, called KS in \cite{Ort_Srini}). The present paper will thus focus on the GKS case when $\mathbf{\Bar{g}}$ is {\em not} of constant curvature.} This enables one to start from more general vacuum seeds, not restricted to the KS form, and thus also obtain more general electrovac solutions.  Furthermore, we will broaden the analysis of \cite{Ort_Srini} in two more directions. First, in odd dimensions we will allow for an additional Chern-Simons (CS) term in the Maxwell equations (not considered in \cite{Ort_Srini}). Moreover, we will study also the case when $\mathbf{k}$ is non-expanding, ultimately leading to a subset of Kundt solutions.

More specifically, our analysis will be based on the following assumptions:

\begin{enumerate}

	\item\label{ass1} The $n$-dimensional line-element and the vector potential take, respectively, the GKS form~\eqref{GKS} and \eqref{aligned_A}, where $\mathbf{k}$ is null, and the electromagnetic field is $\mathbf{F}=d\mathbf{A}$.  
	
	\item\label{ass2} The pair $(\mathbf{g},\mathbf{F}$) is a solution of the Einstein-Maxwell(-CS) theory~\eqref{action}.
		
	\item\label{ass3} The null congruence defined by $\mathbf{k}$ is geodesic and, in addition, satisfies the so-called optical constraint~\eqref{OC} \cite{OrtPraPra09} (see also \cite{type_III_N,Malek:2010mh,OrtPraPra13,OrtPraPra13rev}). We note that these properties can be assumed, equivalently \cite{OrtPraPra09,Malek:2010mh,Ort_Srini,Srinivasan:2025hro}, either w.r.t. $\mathbf{g}$ or $\mathbf{\Bar{g}}$.	
	
	\item\label{ass4} The background metric~$\mathbf{\Bar{g}}$ is Einstein (i.e., solves~\eqref{action} with $\mathbf{F}=0$) and possesses a Weyl aligned null direction (WAND) aligned with $\mathbf{k}$ -- the Weyl type of $\mathbf{\Bar{g}}$ is thus I or more special \cite{Coleyetal04}.	

\end{enumerate}

\begin{remark}

\label{rem_OC}

Assumption~\ref{ass3} can be dropped if one alternatively assumes the Einstein background $\mathbf{\Bar{g}}$ to be itself of the GKS form $\mathbf{\Bar{g}}=\mathbf{\Bar{g}_0} -2H_0\mathbf{k}\otimes \mathbf{k}$, with respect to a different Einstein metric $\mathbf{\Bar{g}_0}$ and the same $\mathbf{k}$ (both $\mathbf{\Bar{g}}$ and $\mathbf{\Bar{g}_0}$ possessing the same cosmological constant). Indeed, in that case it follows automatically that $\mathbf{k}$ is geodesic and obeys the optical constraint \cite{Srinivasan:2025hro} -- see also \cite{OrtPraPra09,Malek:2010mh} in the special case when $\mathbf{\Bar{g}}$ is of constant curvature.\footnote{If $\mathbf{k}$ is non-expanding, it must be Kundt (as follows from the trace of~\eqref{Rij_GKS}, cf. also \cite{OrtPraPra09,Malek:2010mh}), in which case the optical constraint~\eqref{OC} is satisfied trivially.}
(Examples of this kind include the Taub-NUT solutions of Appendix~\ref{appendix_NUT}, while examples {\em not} of this kind will be provided in Section~\ref{subsibsec_5D_example} and Appendix~\ref{subsec_app_Kundt_null}.) In general, the assumption on the optical constraint has a technical character, in that it allows for a significant simplification of the field equations (most notably in the Newman-Penrose formalism \cite{Pravdaetal04,Coleyetal04vsi,OrtPraPra07}). It is also relevant in the context of higher-dimensional formulations of the Goldberg-Sachs theorem \cite{Pravdaetal04,OrtPraPra09,DurRea09,OrtPraPra09b,Ortaggioetal12,OrtPraPra13,OrtPraPra13rev,OrtPraPra18,TinPra19,Taghavi-Chabert22,Srinivasan:2025hro}. 

\end{remark}

\begin{remark}

From assumption~\ref{ass4} it follows that $\mathbf{k}$ is a WAND also of $\mathbf{g}$ (cf.~Propositions~3.8 and 3.9 of \cite{Srinivasan:2025hro}; see also eq.~(8,\cite{Xanthopoulos78}) and \cite{bilge,BilGur86,Nahmad88} in four dimensions).

\end{remark}

\subsection{Summary of results}

The paper starts with setting up the notation and the form of the field equations in the preliminary Section~\ref{Prelims}. Under the above assumptions~\ref{ass1}--\ref{ass4}, our analysis subsequently yields several distinct branches of solutions, which can be summarized as follows.

\begin{enumerate}[a.]
	\item If $\mathbf{k}$ is {\em expanding and twisting}, then its shear must vanish. This can occur only in {\em even} dimensions and the corresponding field $\mathbf{F}=d\mathbf{A}$ is necessarily non-null. If one further specializes assumption~\ref{ass4} to the Weyl type I(a) \cite{Coleyetal04}, all the solutions of this branch are described by \eqref{charge_NUT_ansatz_main}, \eqref{F_soln_NUT} with \eqref{alpha_KS_NUT_main}, \eqref{DF_main}, \eqref{GKS_NUT_vacuum_main}--\eqref{vacu_H_NUT_main}, where the base metric $\mathbf{h}$ is Einstein and (almost-)K\"{a}hler. However, when the Weyl type is strictly I (i.e., not I(a)), we are not able to construct the most general family of solutions, since even the full class of the corresponding vacuum seeds is not known in this case \cite{Taghavi-Chabert22}. This branch is analyzed in Section~\ref{section_charged_GKS_EM_theory}.
	
\item The branch of an {\em expanding and twistfree} $\mathbf{k}$ contains two subacases. (i) If the shear is zero, one obtains a subclass (first obtained in \cite{Tangherlini63,GibWil87}) of the Robinson-Trautman electrovac solutions \cite{OrtPodZof08}, given by \eqref{2_form_charge_RT}, \eqref{F_RT}, where $\mathbf{F}$ is again non-null. Now $\mathbf{k}$ is necessarily an mWAND. (ii) For non-zero shear, instead, the Maxwell field $\mathbf{F}$ must be null. This branch is studied in Section~\ref{non_twisting_section} (with more details in Appendix~\ref{shearing_appendix}). 

\item If $\mathbf{k}$ is {\em non-expanding}, it necessarily possesses vanishing twist and shear, thus defining a Kundt null direction (which is automatically an mWAND \cite{OrtPraPra07,Podolsky:2008ec}). In order to keep it clearly separated from the branches which admit a non-zero expansion, this case is analyzed in Appendix~\ref{app_Kundt}. It is worth observing that this will provide new insight not only in the GKS, but also in the KS case, since the non-expanding case was not considered in \cite{Ort_Srini}.

\end{enumerate}

The remaining appendices are devoted to various auxiliary results. Appendix~\ref{Curvature_appendix} summarizes the form of the Ricci rotation coefficients \cite{Ort_Srini} and of the frame components of the Ricci tensor \cite{Srinivasan:2025hro} for geometries of the GKS type~\eqref{GKS}. Appendix~\ref{appendix_NUT} first reviews basic features of the higher-dimensional vacuum Taub-NUT metrics \cite{Berard82,Bais:1984xb,PagPop87} (cf. also \cite{Lor_Dieter,ChaGib96,Taylor:1998fd,Awad:2000gg,Mann:Nuttier}, and a recent discussion in Appendix~C of \cite{Ort_Srini}), and subsequently provides a derivation of their charged generalization by means of a GKS transformation (thus complementing Section~\ref{section_charged_GKS_EM_theory}). Appendix~\ref{shearing_appendix} presents certain technical details relevant to Section~\ref{subsec_shearing}, i.e., the case when $\mathbf{k}$ is expanding, twistfree and shearing. Additionally, it obtains the full family of solutions in $n=5$ dimensions when $\mathbf{k}$ is assumed to be an mWAND (thus not just a WAND), thus also extending the results of \cite{Ort_Srini} for the special KS subcase.

\section{Preliminaries and field equations}

\label{Prelims}

\subsection{Notation}

The Einstein-Maxwell(-CS) theory in $n$-spacetime dimensions is defined by (cf., e.g., \cite{DesJacTem82,GauMyeTow99})
\begin{align}
    S= &\int d^nx \sqrt{-g}\Big[ \frac{1}{\kappa}(R-2\Lambda) -\frac{1}{2}F^{ab}F_{ab}  \Big] \nonumber\\
    &- 2\chi \sum_{d=1}^{n} \frac{\delta_{n, 2d+1}}{d+1}\int d^{2d+1} x \epsilon^{a_1\dots a_{2d+1}}F_{a_1 a_2} \dots F_{a_{2d-1}a_{2d}}A_{a_{2d+1}} ,
		\label{action}
    \end{align}
where $\kappa$ and $\chi$ are respectively the gravitational and CS couplings, $\epsilon^{a_1 \dots a_{2d+1}}$ is the $(2d+1)$-dimensional constant Levi-Civita symbol, and the CS term contributes only in odd dimensions $n=2d+1$. Variations of~\eqref{action} give rise to the Einstein-Maxwell(-CS) equations
\beqn
 & & R_{ab}-\frac{1}{2}Rg_{ab}+\Lambda g_{ab}=\kappa T_{ab} , \label{Einst} \\ 
 & & \nabla_a{F}^{ab}=\chi j^{b} , \label{Maxw}
\eeqn
where the energy-momentum tensor $T_{ab}$ (not affected by the CS term) and the CS current $j^a$ read
\begin{align}
    &T_{ab}=F_{ac}F_b^{\phantom{b}c}-\frac{1}{4}g_{ab}F_{cd}F^{cd} 
	\label{T},\\
    & j^{a}= \frac{1}{\sqrt{-g}}\sum_{d=1}^{n} \delta_{n, 2d+1} \epsilon^{a a_1a_2\dots a_{2d}}F_{a_1 a_2}\dots F_{a_{2d-1}a_{2d}}.\label{CS_current}
\end{align}

We will adopt the metric signature $(-,+,\dots, +)$ and use the indices as follows: $a,b,\dots=0,1,\dots,n-1$, and $i, j, \dots =2,3,\dots, n-1$. We will work in a null frame \cite{Coleyetal04,Pravdaetal04,OrtPraPra13rev} adapted to the GKS metric form \eqref{GKS} \cite{OrtPraPra09,Malek:2010mh, Ort_Srini, Srinivasan:2025hro}, given by\footnote{With a slight abuse of notation, the symbol $\mathbf{k}$ will denote both the vector field $k^a\pa_a$ and the corresponding covector $k_a\d x^a$ (where $k_a=g_{ab}k^b$). It will be clear from the context what is the object under consideration.}   
\begin{align}
   \{\mathbf{m}_{(0)} = \mathbf{k}, \mathbf{m}_{(1)} = \mathbf{n}, \mathbf{m}_{(i)}\}, \label{frame_GKS}
\end{align}
with $\mathbf{n}$ being a second null vector satisfying $n^{a}k_a = 1$, and $\mathbf{m}_{(i)}$ are $(n-2)$ orthonormal spacelike vectors, i.e., $m_{(i)}^a m^{(j)}_a = \delta_{ij}$, which are also orthogonal to the two null vectors. One can then define the covariant derivatives along these frame directions as follows \cite{Pravdaetal04}
 \be
    D\equiv m^a_{(0)}\nabla_a = k^a \nabla_a, \quad \Delta \equiv m^a_{(1)}\nabla_a =  n^a \nabla_a, \quad \delta_i \equiv  m^{a}_{(i)}\nabla_a,\label{Frame_derivatives_GKS}\\
\ee
 and thereby also the associated Ricci rotation coefficients
 \be
  L_{ab}= m^{c}_{(a)}m^{d}_{(b)} k_{c;d}, \quad  N_{ab}= m^{c}_{(a)}m^{d}_{(b)} n_{c;d}, \quad  \overset{i}{M}_{ab}= m^{c}_{(a)}m^{d}_{(b)} m^{(i)}_{c;d}, \label{GKS_Ricci_rot}
\ee
which satisfy the following relations
\begin{align}
  L_{0a}=N_{1a}=N_{0a}+L_{1a}=\M{i}{0}{a} + L_{ia} = \M{i}{1}{a}+N_{ia}=\M{i}{j}{a}+\M{j}{i}{a}=0.  \label{Ricci_rot_identities}
\end{align}
Eq.~\eqref{GKS_Ricci_rot} implies $k^a k_{b;a}= L_{10}k_b+L_{i0}m^{(i)}_{b}$, i.e., $\mathbf{k}$ is geodesic iff $L_{i0}=0$, and affinely parametrized if, additionally, $L_{10}=0$ \cite{Pravdaetal04}.

The optical matrix of $\mathbf{k}$, defined as the spatial projection of its covariant derivative, is given by the Ricci rotation coefficient $L_{ij}$ (cf.~\eqref{GKS_Ricci_rot}), and defines the following optical quantities \cite{Pravdaetal04, OrtPraPra07, OrtPraPra09,OrtPraPra13rev}
\begin{align}
&S_{ij}\equiv L_{(ij)}, \quad\theta \equiv \frac{1}{n-2} S_{ii}, \quad   \sigma_{ij} \equiv S_{ij}- \theta \delta_{ij}, \label{optical_decompose}\\
& \quad \sigma^2 \equiv \sigma_{ij}\sigma_{ij},  \quad A_{ij}\equiv L_{[ij]}, \quad \omega^2 \equiv A_{ij}A_{ij}, \label{optical_scalars}
\end{align}
where $\theta$, $\sigma^2$, and $\omega^2$ are the optical scalars, representing the expansion, shear, and twist of $\mathbf{k}$, respectively.

Analogous to \eqref{frame_GKS}, one can define a null frame for the background, $\bar{\mathbf{g}}=\mathbf{g}|_{H=0}$, and hence also the frame-projected derivatives and the Ricci rotation coefficients (see Appendix \ref{Curvature_appendix}). We will use bars to distinguish the quantities defined on the background from their counterparts in the full geometry.

As noted in \cite{Srinivasan:2025hro} (cf. also~\eqref{Li0_GKS}, and \cite{OrtPraPra09,Malek:2010mh,Ort_Srini} for earlier results in the KS case, and Section~\ref{sec_intro} for references in $n=4$), the optical matrix, and hence the optical scalars of $\mathbf{k}$, are invariant under a GKS transformation. As already stated (cf. assumption~\ref{ass4}), we assume that $\mathbf{k}$ is a geodesic WAND of the background, i.e., $\bar L_{i0}=0=\bar C_{0i0j}$. From the results of \cite{Srinivasan:2025hro} (in particular, Propositions~3.8 and 3.9), it follows that these properties are also inherited by $\mathbf{k}$ in the full geometry, i.e., $L_{i0}=0=C_{0i0j}$. Furthermore, without loss of generality, we can take $\mathbf{k}$ to be affinely parametrized in the full geometry (and hence also in the background; see equation \eqref{Li0_GKS}) and denote the affine parameter by $r$, so that $k^a\partial_a=\partial_r$. In addition, we take also the other frame vectors to be parallelly transported along $\mathbf{k}$ \cite{OrtPraPra07}, such that, altogether, 
\begin{align}
    L_{i0}=0=L_{10} ,  \qquad \overset{i}{M}_{j0}=0=N_{i0}.
		\label{parallel_transp}
\end{align}

The background being Einstein implies $\bar R_{ab}k^ak^b=0$. Thanks to the geodesicity of $\mathbf{k}$, it follows that also $R_{ab}k^ak^b=0$ (cf. Proposition~2.1 of \cite{Srinivasan:2025hro}).\footnote{Upon using the Einstein equations, this means $\bar T_{ab}k^ak^b=0=T_{ab}k^ak^b$. It will follow from~\eqref{F_compts} that $T_{ab}k^a k^b=0$ is indeed identically satisfied by the Maxwell(-CS) energy-momentum tensor corresponding to the vector potential~\eqref{aligned_A}.} Combined with the fact that (by assumption) $\mathbf{k}$ is a WAND, this means that it is also a Riemann AND both of the full and of the background geometry, i.e., 
\begin{align}
    \bar R_{0i0j}=0=R_{0i0j}. \label{Riemann_AND}
\end{align}

For later purposes, let us list the Ricci identities $(11b)$, $(11e)$, $(11g)$, and $(11k)$ of \cite{OrtPraPra07}, which, using \eqref{parallel_transp} and \eqref{Riemann_AND}, reduce to
\begin{align}
  &D L_{1i}=   -L_{1j}L_{ji}- R_{010i},\label{Ricci_1}\\
  & D L_{i1}=- L_{ij}L_{j1}- R_{010i},\label{Ricci_2}\\
  & D L_{ij}= -L_{ik}L_{kj},\label{Ricci_3}\\
  & \delta_{[j|}L_{i|k]}= L_{1[j|}L_{i|k]}+ L_{i1}A_{jk}+L_{il}\overset{l}{M}_{[jk]}+ L_{l[j|}\overset{l}{M}_{i|k]}- \frac{1}{2}R_{0ijk}.\label{Ricci_4}
\end{align}

Thanks to~\eqref{parallel_transp}, the derivative commutation relations \cite{Coleyetal04vsi} read 
\beqn
 & & \T D - D \T=L_{11} D + L_{i1} \delta_i , \label{comm_DelD} \\
 & & \delta_i D-D\delta_i=L_{1i}D+L_{ji}\delta_j , \label{comm_dD} \\
 & & \delta_i \T - \T \delta_i  = N_{i1} D + (L_{i1}-L_{1i}) \T + (N_{ji}+\M{j}{i}{1}) \delta_j , \label{comm_dDel} \\
 & & \delta_{[i}\delta_{j]}=N_{[ij]}D+L_{[ij]}\Delta+\M{k}{[i}{j]}\delta_k . \label{comm_dd}
\eeqn

\subsection{Structure of the optical matrix}

\label{struc_opt_mat}

Throughout the paper, it will be assumed that the optical matrix of $\mathbf{k}$ satisfies the optical constraint (cf. assumption~\ref{ass3}) \cite{OrtPraPra09,type_III_N,Malek:2010mh,OrtPraPra13,OrtPraPra13rev}, i.e., 
\begin{align}
     L_{ik}L_{jk}\propto S_{ij} . \label{OC}
\end{align}
For the role played by this condition in the context of a higher-dimensional formulation of the Goldberg-Sachs theorem we refer the reader to \cite{Pravdaetal04,OrtPraPra09,OrtPraPra09b,Ortaggioetal12,OrtPraPra13,OrtPraPra13rev,OrtPraPra18,TinPra19,Srinivasan:2025hro}.

\subsubsection{Non-expanding case ($\theta=0$)}

\label{subsubsec_nonexp}

Thanks to $\theta=0$, the optical constraint~\eqref{OC} implies $\omega^2=0=\sigma^2$, i.e, $L_{ij}=0$ and $\mathbf{k}$ is a Kundt congruence. The analysis of this case will be relegated to Appendix~\ref{app_Kundt}.

\subsubsection{Expanding case ($\theta\neq0$)}

When $\theta\neq0$, one can use the trace of~\eqref{OC} to rewrite it as \cite{OrtPraPra09}
\begin{align}
     L_{ik}L_{jk}= \frac{L_{lk}L_{lk}}{(n-2)\theta} S_{ij}, \label{opt_constraint}
\end{align}

As noted in \cite{type_III_N,Ort_Srini,Srinivasan:2025hro}, the optical constraint implies that the results of \cite{OrtPraPra09,Malek:2010mh} can be directly applied to block diagonalize the optical matrix of the KS vector into the following form \cite{OrtPraPra09,type_III_N,Malek:2010mh,OrtPraPra13,OrtPraPra13rev}
\beqn
L_{ij}=\left(\begin {array}{cccc} \fbox{${\cal L}_{(1)}$} & & &  \\
& \ddots & & \\ 
& & \fbox{${\cal L}_{(p)}$} & \label{L_general} \\
& & & \fbox{$\begin {array}{ccc} & & \\ \ \ & \tilde{\cal L} \ \ & \\ & & \end {array}$}
\end {array}
\right) . \label{L_ij_canonical}
\eeqn

The blocks ${\cal L}_{(1)}, \dots, {\cal L}_{(p)}$ are $2\times 2$ matrices, while $\tilde{\cal L}$ is a diagonal matrix of size $(n-2-2p)\times(n-2-2p)$, and they are given by
\beqn
& & {\cal L}_{(\mu)}=\left(\begin {array}{cc} s_{(2\mu)} & A_{2\mu,2\mu+1} \nonumber \\
-A_{2\mu,2\mu+1} & s_{(2\mu)} 
\end {array}
\right) \qquad (\mu=1,\ldots, p) , \\
& & s_{(2\mu)}=\frac{r}{r^2+(a^0_{(2\mu)})^2} , \qquad A_{2\mu,2\mu+1}=\frac{a^0_{(2\mu)}}{r^2+(a^0_{(2\mu)})^2} , \label{s_A} \\
& &  \tilde{\cal L}=\frac{1}{r}\mbox{diag}(\underbrace{1,\ldots,1}_{(m-2p)},\underbrace{0,\ldots,0}_{(n-2-m)}) , \label{diag_block_Lij_canonical}
\eeqn
where $a^0_{(2\mu)}\neq0$ are $r$-independent integration functions, and $m$ is the rank of $L_{ij}$; hence $n-2\geq m\geq 2p \geq 0$. From the above block structure, one obtains the following expressions for the optical scalars \cite{OrtPraPra09,type_III_N}
\beqn
 & & (n-2)\theta=2\sum_{\mu=1}^p\frac{r}{r^2+(a^0_{(2\mu)})^2}+\frac{m-2p}{r} , \label{exp} \\
 & & \omega^2=2\sum_{\mu=1}^p\left(\frac{a^0_{(2\mu)}}{r^2+(a^0_{(2\mu)})^2}\right)^2 , \label{twist} \\
 & & \sigma^2=2\sum_{\mu=1}^p\left(\frac{r}{r^2+(a^0_{(2\mu)})^2}-\theta\right)^2+(m-2p)\left(\frac{1}{r}-\theta\right)^2+(n-2-m)\theta^2 . \label{shear} 
\eeqn
Observe that $\theta\neq 0$ implies $m>0$, and $\omega\neq 0 \Leftrightarrow p>0$. For the case of expanding $\mathbf{k}$, there are two possible ways in which the shear can vanish \cite{OrtPraPra09, Ort_Srini,Srinivasan:2025hro}: (i) $p\neq0$,  $m=2p=n-2$, $(a^0_{(2)})^2=\dots= (a^0_{(2p)})^2$ or (ii) $p=0$, $\theta=\frac{1}{r}$, $m=n-2$.

\subsection{Maxwell(-CS) equations}

\label{subsection_Maxwell_eqn}

Using~\eqref{aligned_A} and \eqref{parallel_transp}, one finds that the non-zero components of $\mathbf{F}$ are given, as in \cite{Ort_Srini}, by
\be
    F_{01}=D\alpha  , \qquad F_{ij}= -2\alpha A_{ij} , \qquad F_{1i}=-2\alpha L_{[1i]} - \delta_{i}\alpha  .
	\label{F_compts}		
\ee
It follows that $\mathbf{F}$ is aligned with $\mathbf{k}$ (i.e., $F_{0i}=0$) and it is easy to check that for the CS current \eqref{CS_current}, the $j^1$ component vanishes, whereas the components $j^0, j^i$ can be non-zero when $F_{ij}\neq0$. However, we will not require their precise forms at this stage. Therefore, the Maxwell(-CS) equations \eqref{Maxw} take the following form \cite{Durkeeetal10, Ortaggio:2014ipa,Ort_Srini}\footnote{In fact, eqs.~\eqref{Max1}--\eqref{Max2} are not specific to the GKS assumption and hold true, more generally, for any Maxwell(-CS) field such that~\eqref{aligned_A} and \eqref{parallel_transp} are satisfied (i.e., $\mathbf{k}$ can be any geodesic, affinely parametrized, null vector -- not assumed to be a WAND).
In particular, when $j^a=0$, eqs.~\eqref{Max1}--\eqref{Max2} are identical to the Maxwell equations obtained for KS spacetimes in \cite{Ort_Srini}. Let us further note that, similar to \cite{Ort_Srini}, a $DL_{[1i]}$ term was eliminated from \eqref{Max3} by using the Ricci identities \eqref{Ricci_1}, \eqref{Ricci_2}.}
\beqn
  & & D^2\alpha + (n-2)\theta D\alpha +2\alpha \omega^2=0	,		\label{Max1} \\
  & &  \left[D\delta_i+(3A_{ji}-\sigma_{ij})\delta_j +(n-3)\theta\delta_{i}  +(L_{1i}-2L_{i1})D\right]\alpha  \nonumber \\
	& & \qquad\qquad {}+2 \alpha\Big[\delta_{j}A_{ji} +(n-4)\theta L_{[1i]}-2\sigma_{ij}L_{[1j]}-2A_{ji}L_{j1}+ \overset{k}{M}_{jj}A_{ki} +\overset{k}{M}_{ij}A_{jk}\Big]=-\chi j^i ,
 		\label{Max3} \\
  & &   (\Delta D + \delta_j \delta_j+4 L_{[1j]}\delta_{j}+ N_{jj}D + \overset{k}{M}_{jj}\delta_{k}) \alpha  \nonumber \\
  & & \qquad\qquad  {}+2\alpha(2L_{[1j]}L_{[1j]}+\delta_{j}L_{[1j]}+ \overset{k}{M}_{jj}L_{[1k]} +N_{kj}A_{jk}) =\chi j^0.
				\label{Max2}  		
\eeqn

Since the non-expanding case will be studied in Appendix~\ref{app_Kundt}, we give below the general solutions to \eqref{Max1} for the expanding case (which can be either twisting or twistfree), thus fixing the $r$-dependence of the Maxwell(-CS) potential $\alpha$.

\subsubsection{Expanding, twisting $\mathbf{k}$}

The most general solutions to \eqref{Max1} for the case of expanding $\mathbf{k}$ were already obtained in \cite{Ort_Srini} by making use of the structure of the optical matrix \eqref{L_ij_canonical}--\eqref{diag_block_Lij_canonical} and the consequent expressions for the optical scalars \eqref{exp}--\eqref{shear}, and we state them directly. When $\mathbf{k}$ is expanding and twisting ($p\ge1\Rightarrow m\ge2$), the $r$-dependence of $\alpha$ is given by \cite{Ort_Srini}
\be
    \alpha = \frac{\beta}{r^{m-2p-1}}\prod _{\mu=1}^{p} \frac{1}{r^2 + (a^0_{2\mu})^2}
		 \label{rescale_alpha}.
\ee
The auxiliary function $\beta$ above takes the following forms, depending on whether $m$ is odd or even
\beqn
   & & \beta= \alpha_{0}+  \beta_{0}\sum_{\mu=0}^p \frac{{\cal A}^0_{\mu}}{m-1-2\mu}r^{m-1-2\mu} \qquad (m\ge 2 \mbox{ even}) , \label{beta_even} \\
	 & & \beta= \alpha_{0}+  \beta_{0}\Bigg({\cal A}^0_p\delta_{m,2p+1}\ln r+ \sum_{\substack{\mu=0 \\
    (2\mu\neq m-1)}}^p \frac{{\cal A}^0_{\mu}}{m-1-2\mu}r^{m-1-2\mu}\Bigg) \qquad (m\ge 3 \mbox{ odd}) ,
\eeqn
where
\be
 {\cal A}^0_{0}=1 , \qquad\quad {\cal A}^0_{\mu}=\sum_{\nu_1<\nu_2<\ldots<\nu_\mu} (a^{0}_{(2\nu_1)})^2(a^{0}_{(2\nu_2)})^2\ldots(a^{0}_{(2\nu_{\mu})})^2	\quad (\mu=1,\ldots, p) ,
 \label{A_alpha}
\ee
with $\alpha_{0}$ and $\beta_{0}$ being $r$-independent integration functions. 

\subsubsection{Expanding, twistfree $\mathbf{k}$}

When $\mathbf{k}$ has a vanishing twist ($p=0$), the $r$-dependence of $\alpha$ is fixed as \cite{Ort_Srini}
\beqn
	 & & \alpha = \alpha_{0}r^{1-m}+  \frac{\beta_{0}}{m-1} \qquad (m\neq1) ,  \label{alpha_nontwist} \\
	 & & \alpha = \alpha_{0}+\beta_{0}\ln r  \qquad (m=1) . \label{alpha_nontwist_m=1}
\eeqn

\subsection{Einstein equations}
\label{subs_Einstein_eqns}

Since the Ricci tensor of the full geometry is naturally expressed in terms of the Ricci tensor of the background (Appendix~\ref{Curvature_appendix}), it is convenient to rewrite the Einstein equations~\eqref{Einst} with~\eqref{T} as
\begin{align}
    R_{ab}=\frac{2\Lambda}{n-2}g_{ab}+\kappa \left[F_{ac}F_b^{\phantom{b}c}-\frac{1}{2(n-2)}g_{ab}F_{cd}F^{cd}\right].
		\label{Eint_full_geo}
\end{align}
Using~\eqref{F_compts}, the non-vanishing frame components of \eqref{Eint_full_geo} take the form\footnote{The explicit form of $T_{ab}$ following from~\eqref{F_compts} can be found in \cite{Ort_Srini}.}
\begin{align}
    & R_{01}-\frac{2\Lambda}{n-2}=-\kappa\left[\frac{n-3}{n-2}(D\alpha)^2+ \frac{2\alpha^2\omega^2}{n-2}\right], \label{01_eqn}\\
    & R_{ij}-\frac{2\Lambda}{n-2}\delta_{ij}=\kappa\left[4\alpha^2 A_{ik}A_{jk} + \frac{\delta_{ij}}{n-2}\left[ (D\alpha)^2-2\alpha^2\omega^2\right]\right], \label{ij_eqn}\\
    & R_{1i}=-\kappa\left[(D\alpha) (2\alpha L_{[1i]}+ \delta_i \alpha) - 2\alpha A_{ij}(2\alpha L_{[1j]}+\delta_{j}\alpha)\right], \label{1i_eqn}\\
    & R_{11}=\kappa(2\alpha L_{[1i]} + \delta_{i}\alpha)(2\alpha L_{[1i]} + \delta_{i}\alpha),\label{11_eqn}
\end{align}

Discussing equations \eqref{1i_eqn} and \eqref{11_eqn} would not be illuminating at this stage. Therefore, we focus on those that fix the $r$-dependence, namely \eqref{01_eqn} and \eqref{ij_eqn}, below and in Sections~\ref{section_charged_GKS_EM_theory} and \ref{non_twisting_section} for the expanding case, and in Appendix~\ref{app_Kundt} when $\theta=0$.

\subsubsection{Expanding $\mathbf{k}$}

By construction (cf. assumption~\ref{ass4} in Section~\ref{sec_intro}), the background is Einstein with the same value of $\Lambda$ as the full space, i.e., $\bar R_{ab}=\frac{2\Lambda}{n-2}\bar g_{ab}$.
Thanks to~\eqref{Rij_GKS} with $\bar R_{ij}=\frac{2\Lambda}{n-2}\delta_{ij}$, \eqref{Riemann_AND} and \eqref{opt_constraint}, one can decompose~\eqref{ij_eqn} into the following (spatial) trace and tracefree parts 
\beqn 
		& & 2DH + 2H(n-2)\theta  - 2H\frac{L_{mn}L_{mn}}{(n-2)\theta} = -\frac{\kappa}{(n-2)\theta} \left[ 2\alpha^2 \omega^2 + (D\alpha)^2\right] , \label{Eii} \\ 
    & & S_{ij}\frac{(D\alpha)^2+2\alpha^2 \omega^2}{(n-2)\theta}= 4 \alpha^2 A_{ik}A_{jk} + \frac{\delta_{ij}}{n-2}\left[(D\alpha)^2 -2\alpha^2 \omega^2\right] . \label{Eij_tracefree} 
\eeqn

Equations \eqref{Eii} and \eqref{Eij_tracefree} take the same form as in \cite{Ort_Srini}. Note also that \eqref{Eij_tracefree} is identically satisfied in the shearfree case thanks to the Sachs equations \cite{OrtPraPra07}. The block structure~\eqref{L_ij_canonical}--\eqref{diag_block_Lij_canonical} of the optical matrix thus constrains the matter field $\alpha$ and the optical scalars through~\eqref{Eij_tracefree}, giving rise to multiple conditions depending on the values of $m$ and $p$ \cite{Ort_Srini}. For the twisting case, i.e., $p\ge1$, each of the $\mathcal{L}_{(\mu)}$ block leads to the respective $\mu$-dependent condition given below
\be
    \frac{r}{r^2 + (a^{0}_{(2\mu)})^2}\frac{(D\alpha)^2+2\alpha^2 \omega^2}{(n-2)\theta}=4 \alpha^2 \left(\frac{a^{0}_{(2\mu)}}{r^2 + (a^{0}_{(2\mu)})^2}\right)^2 + \frac{1}{n-2}\left[(D\alpha)^2 -2\alpha^2 \omega^2\right] \qquad (p\ge1) .
\label{Eij_block}		
\ee
The cases $m>2p$ and $m<n-2$, respectively, imply the following constraints arising from the non-zero and zero entries of the $\tilde {\mathcal L}$ block
\beqn
    & & (D\alpha)^2 -2\alpha^2 \omega^2=\frac{1}{r\theta}\left[(D\alpha)^2+2\alpha^2 \omega^2\right] \qquad\quad (m>2p) , \label{Eij_1} \\
		& & (D\alpha)^2 -2\alpha^2 \omega^2=0 \qquad\quad (m<n-2) . \label{Eij_0}
\eeqn

Equation \eqref{01_eqn} is equivalent to a linear combination of~\eqref{Eii}, its $D$-derivative, and \eqref{Max1}, upon using~$\bar R_{01}=\frac{2\Lambda}{n-2}$ (recall~\eqref{R01_GKS}), \eqref{Ricci_3}, and the contraction of \eqref{Eij_tracefree} with $\sigma_{ij}$; hence, it can be ignored hereafter.\footnote{In \cite{Ort_Srini} we incorrectly claimed this follows from the Bianchi identity -- the results of \cite{Ort_Srini} remain, however, correct.}

\section{Twisting solutions} 

\label{section_charged_GKS_EM_theory}

\subsection{Shearfree condition}

For the special case when the background, $\mathbf{\bar g}$, is an Einstein-KS spacetime and the dynamics is governed by the Einstein-Maxwell theory, it was shown in \cite{Ort_Srini} that if $\mathbf{k}$ is expanding and twisting, then it can be consistent with equations \eqref{Max1} (i.e., \eqref{rescale_alpha}--\eqref{A_alpha}), \eqref{Eij_tracefree}--\eqref{Eij_0} with \eqref{exp}--\eqref{shear} only if it is shearfree, and thus for $m=2p=n-2$, with $n$ even (cf. Section~\ref{struc_opt_mat}). Since those equations hold unchanged in the present (more general) case, the same argument as in \cite{Ort_Srini} leads to 
\begin{proposition}\label{prop_charging_GKS}
   Let $\mathbf{\bar g}$ be an Einstein spacetime in $n\geq4$, with a geodesic, expanding, twisting WAND $\mathbf{k}$ satisfying the optical constraint \eqref{opt_constraint}. Then, $\mathbf{\bar g}$ can be charged in the Einstein-Maxwell(-CS) theory by a GKS transformation \eqref{GKS} with \eqref{aligned_A} only if $n$ is even and $\mathbf{k}$ is shearfree (thus the Einstein equation~\eqref{Eij_tracefree} becomes an identity and the CS term~\eqref{CS_current} vanishes).
\end{proposition}

It thus appears that the first step in the charging procedure consists in selecting a vacuum background admitting a {\em shearfree} WAND $\mathbf{k}$.  All $n>4$ Einstein spacetimes possessing a geodesic, expanding, twisting, shearfree congruence $\mathbf{k}$ are known \cite{Taghavi-Chabert22} if, additionally, one assumes the condition $k_{[e} C_{a]bcd} k^b k^c = 0$, i.e. (cf.~\cite{Ortaggio09}), the Weyl type to be I(a) or more special, which implies that $\mathbf{k}$ is a WAND obeying the optical constraint (cf. Section~\ref{subsec_NUT_vac}; in that case, a second WAND with the same properties also exists \cite{Taghavi-Chabert22,Ort_Srini}, i.e., the Weyl type automatically becomes I$_i$(a) or more special). Under the stronger condition $k_{[e} C_{a]bcd} k^b k^c = 0$ (Weyl type I(a) or more special), we are able to obtain below the complete family of $n>4$ solutions using the charging procedure of Proposition~\ref{prop_charging_GKS}. For the general case of a shearfree $\mathbf{k}$ (i.e., without the additional assumption $k_{[e} C_{a]bcd} k^b k^c = 0$), the needed vacuum seeds are not known and therefore their charged counterparts cannot be explicitly found either. Before proceeding with the type I(a) case, let us make two remarks on the result found above.

\begin{remark}

\label{rem_CCLP}

The five-dimensional charged rotating Chong-Cveti{\v c}-L{\" u}-Pope (CCLP) solution \cite{CCLP}, which also belongs to the GKS class \cite{Srinivasan:2025hro, Hassaine:2024mfs} with $\mathbf{k}$ being a WAND obeying assumption~\ref{ass3} \cite{Malek_xKS}, does not contradict Proposition~\ref{prop_charging_GKS}, as the background of the GKS form of the CCLP metric is not an Einstein spacetime, but is coupled to the Maxwell-CS matter field.

\end{remark}

\begin{remark}

Electrically charged four-dimensional (A)dS-Kerr-NUT metrics \cite{Carter68pla,Carter68cmp} form an example of charged GKS spacetimes \cite{Hassaine:2024mfs} consistent with Proposition~\ref{prop_charging_GKS}, as they possess a shearfree KS vector in accordance with the Goldberg-Sachs theorem \cite{Gold_Sachs_62}. By contrast, for $n>4$, neither the multi-NUT solutions \cite{Mann:Nuttier,LuPagPop04,multi_NUT_Mann} nor the (A)dS-Kerr-NUT metrics \cite{Chen_Pope_Kerr_NUT_multi} can be charged by the current charging procedure, since their $\mathbf{k}$ is shearing \cite{Pravda_type_D,OrtPraPra13,Ortaggio17,Taghavi-Chabert22} (that the metrics of \cite{Mann:Nuttier,LuPagPop04,multi_NUT_Mann} and \cite{Chen_Pope_Kerr_NUT_multi} satisfy the assumptions of Proposition~\ref{prop_charging_GKS} follows from \cite{Srinivasan:2025hro,Srinivasan_PhD} (see also \cite{OrtPraPra13} in a special case) and \cite{Pravda_type_D}, respectively). Therefore, Proposition \ref{prop_charging_GKS} forms a ``no-go'' result for the charging of these higher-dimensional metrics. For the special case of multiply rotating Myers-Perry metrics \cite{MyePer86} and their extensions to an arbitrary cosmological constant \cite{Gibbonsetal05,Gibbons_rot_HD_cosmol}, such a ``no-go'' result was already established in \cite{Ort_Srini} (cf. also \cite{MyePer86} for an earlier version of the result with a single rotation and $\Lambda=0$). However, as we shall see in the following subsections, charged (A)dS-Taub-NUT solutions in $n>4$  \cite{ManSte06,Awad06} can be understood as charged GKS solutions generated by the charging procedure of Proposition \ref{prop_charging_GKS}, since they do possess a shearfree KS vector \cite{Ortaggio17,OrtPraPra13,Taghavi-Chabert22}.\footnote{Throughout the paper, by ``(AdS-)Taub-NUT'' metrics we will mean the class of geometries reviewed in Appendix~\ref{appendix_NUT} (where relevant references can also be found). From the viewpoint of \cite{Mann:Nuttier,multi_NUT_Mann}, these can be understood as multi-NUT metrics for which all the NUT parameters are equal.} In fact, as will be argued in Section~\ref{subsec_k_mWAND} using the results of \cite{Taghavi-Chabert22}, these are the only possible solutions when $\mathbf{k}$ is an mWAND (cf. also \cite{Ort_Srini}).

\end{remark}

\subsection{Type~I(a) or more special: the complete family of solutions}

\label{subsec_k_mWAND}

Let us assume the WAND $\mathbf{k}$ of the vacuum seed $\mathbf{\bar g}$ to obey $k_{[e} C_{a]bcd} k^b k^c = 0$ (implying the Weyl type~I(a) or more special). By~Proposition~\ref{prop_charging_GKS} and the comments following it, to obtain the complete $n>4$ family of solutions generated by the charging procedure described in Section~\ref{sec_intro}, one must start with the most general family of  vacuum seeds possessing an expanding, twisting, geodesic WAND $\mathbf{k}$   which is, additionally, {\em shear-free} (in addition to the assumption $k_{[e} C_{a]bcd} k^b k^c = 0$). The results of \cite{Taghavi-Chabert22} imply that these solutions coincide with the higher-dimensional Taub-NUT metrics of \cite{Berard82,Bais:1984xb,PagPop87} (see also \cite{Lor_Dieter,ChaGib96,Taylor:1998fd,Awad:2000gg,Mann:Nuttier}), cf. a discussion in \cite{Ort_Srini}. The form of these vacua most useful for our purposes, adapted to the WAND $\mathbf{k}$, is \cite{Ort_Srini} (see also Appendix~\ref{appendix_NUT}, in particular \eqref{GKS_NUT_vacuum} with \eqref{vacu_H_NUT}, and references therein)
\begin{align}
   \mathbf{g}_{\small NUT}=&d r\otimes\mathbf{k}+\mathbf{k}\otimes d r+(r^2+l^2)\mathbf{h} -2\H(r)\mathbf{k}\otimes\mathbf{k},\label{GKS_NUT_vacuum_main}\\
   \mathbf{k}=&d u-2\mathbf{Z},\label{NUT_KS_vector}
\end{align}
where the metric $\mathbf{h}$ on the $(n-2)$-dimensional base space is (almost-)K\"{a}hler-Einstein, $\mathbf{Z}=Z_i dx^i$ is a $1$-form on $\mathbf{h}$ such that $\cF \equiv \d\bZ$ defines the corresponding fundamental $2$-form, and $l$ is the NUT parameter~\eqref{F^2_condition}. The function $\mathcal{H}$ reads 
\begin{align}
  -2\mathcal{H}=&\frac{r}{(r^2+l^2)^{\frac{n-2}{2}}}\left[\mu_0+ \frac{1}{n-2}\int^r d\bar r \Big[\Big(2\Lambda-\frac{\tilde R}{\bar r^2+l^2}\Big)\frac{(\bar r^2+l^2)^{\frac{n}{2}}}{\bar r^2}\Big] \right], 
	\label{vacu_H_NUT_main} 
\end{align}
where $\mu_0$ is an integration constant related to the mass, and $\tilde R$ is the (constant) Ricci scalar of the base space. 
 After going through the steps of charging given in Appendix~\ref{charge_NUT_appnedix}, one obtains the electrovac solution 
\begin{align}
    &\mathbf{g}= \mathbf{g}_{\small NUT} -\frac{\kappa r Q(r)}{(r^2+l^2)^{\frac{n-2}{2}}} \mathbf{k}\otimes \mathbf{k},\label{charge_NUT_ansatz_main}\\
    & \mathbf{F}=(D\alpha)d r \wedge\mathbf k-2\alpha\cF,\label{F_soln_NUT}
\end{align}
with the functions $\alpha(r)$ and $Q(r)$ given by
\begin{align}
&\alpha =\frac{r}{(r^2+l^2)^{\frac{n-2}{2}}} \Bigg[\alpha_{0}+  \beta_{0}\sum_{\mu=0}^{\frac{n-2}{2}} \binom{\frac{n-2}{2}}{\mu} \frac{l^{2\mu}}{n-3-2\mu}r^{n-3-2\mu}\Bigg] ,
	\label{alpha_KS_NUT_main}	\\
    &DQ=-r^{-2}(r^2+l^2)^{\frac{n-4}{2}}\left[2\alpha^2 l^2 +\textstyle{\frac{1}{n-2}}(r^2+l^2)^2(D\alpha)^2\right] ,\label{DF_main}
\end{align}
where $\alpha_0$ and $\beta_0$ are integration constants related to electric charge and asymptotic magnetic field strengths, respectively (cf. \cite{FloQue19,Ort_Srini}). The two WANDs are defined by $\mathbf{k}$ and $\mathbf{n}=dr-(\mathcal{H}+H)\mathbf{k}$ with~\eqref{H_Q} (no other WANDs are possible, cf. Appendix~\ref{charge_NUT_appnedix} and \cite{Srinivasan_PhD}).

There are thus two distinct subfamilies of solutions. (i) If $\mathbf{h}$ is (Einstein and) strictly almost-K\"{a}hler (and thus the covariant derivative of $\cF$  w.r.t. the base space does not vanish), then $\mathbf{k}$ and $\mathbf{n}$ are single WANDs and the Weyl type is I(a). (ii) By contrast, if $\mathbf{h}$ is (Einstein and) K\"{a}hler, then both $\mathbf{k}$ and $\mathbf{n}$ are double WANDs and the Weyl type is D (cf. Appendix~C of \cite{Ort_Srini} and \cite{Pravda_type_D}). The case~(ii) solutions can be identified as the higher-dimensional charged Taub-NUT spacetimes, first obtained in \cite{ManSte06} for the special case $\beta_0=0$, and in full generality in \cite{Awad06} (see also \cite{DehKoh06}). In both cases, the optical matrix of both $\mathbf{k}$ and $\mathbf{n}$ obeys the optical constraint (cf. Section~\ref{appendix_NUT} for more details). In the limit $n=4$, only branch~(ii) survives and one recovers certain solutions first found in \cite{Brill64,Carter68pla,Carter68cmp,Ruban72}.

Recall that when the spacetime is required to be KS \cite{Ort_Srini} (and not just GKS), the base space $\mathbf{h}$ in~\eqref{GKS_NUT_vacuum_main} is further constrained to be (an Einstein-K\"{a}hler space) of constant holomorphic sectional curvature \cite{Bochner1947CurvatureIH,yano1965differential,Okubo_1970}, with the fine-tuning $(n-1)\tilde R=2n\Lambda l^2$, which also simplifies the form of~\eqref{vacu_H_NUT_main} \cite{Ort_Srini}.

\subsection{Explicit examples in $n=6$}

\label{6d_examples_aligned_A}

For explicit examples of charged Taub-NUT, let us take the base space to be the four-dimensional Ricci-flat, Riemannian metric constructed in \cite{1999CQGra..16L...9N} (cf. also \cite{Armstrong2002AnAF} for related discussions), which was shown to admit both a strictly almost-K\"{a}hler as well as a K\"{a}hler structure. The metric in real coordinates is given by \cite{1999CQGra..16L...9N}
\begin{align}
    \mathbf{h}=x(dx^2+ dy^2+ dz^2) + \frac{1}{x}\left(\frac{1}{2}zdy-\frac{1}{2}ydz+dw\right)^2.\label{example_base}
    \end{align}
We will use its strictly almost K\"{a}hler structure to construct an example of charged Taub-NUT with a single WAND $\mathbf{k}$, and likewise its K\"{a}hler structure for the case where $\mathbf{k}$ is a double WAND. However, in both cases, the Maxwell field is given by~\eqref{F_soln_NUT}, the metric takes the form~\eqref{charge_NUT_ansatz_main} with \eqref{GKS_NUT_vacuum_main}, \eqref{NUT_KS_vector} and \eqref{example_base}, and the functions $\mathcal{H}$, $\alpha$, and $Q(r)$ are given by (see~\eqref{vacu_H_NUT_main} with $\tilde R=0$, \eqref{alpha_KS_NUT_main}, \eqref{DF_main})
\begin{align}
    &-2\mathcal{H}=\frac{\mu_0 {r}}{(r^2+l^2)^2}+\frac{\Lambda}{10}\frac{r^6+5 l^2 r^4+15l^4 r^2-5 l^6}{(l^2+r^2)^2}, \label{Hvac_6D} \\
    & \alpha= \frac{r^4}{(r^2+l^2)^2}\Big[ \frac{\alpha_0}{r^3}+ \beta_0 \Big ( \frac{1}{3}+ \frac{2l^2}{r^2}-\frac{l^4}{r^4}\Big)\Big],\label{alpha_soln_6d}\\
    & Q=\frac{9\alpha_0^2(3 r^2+l^2)+96\alpha_0\beta_0l^2 r^3+8\beta_0^2 l^2(-r^6+15 l^2r^4+9l^4 r^2+9l^6)}{36r(r^2+l^2)^2}.\label{Q_soln_6d}
\end{align}

\subsubsection{Example with a single WAND $\mathbf{k}$ (Weyl type~I(a))}

The closed 2-form corresponding to the strictly alomost K\"{a}hler structure of metric~\eqref{example_base} is given by \cite{1999CQGra..16L...9N}
\begin{align}
   \cF = l(x dx\wedge dz +\frac{y}{2}dy \wedge dz-dy\wedge dw), \label{almost_Kahler_example}
\end{align}
which can be obtained from the following $1$-form
\begin{align}
    \mathbf{Z}= -l(xzdx+\frac{1}{2}yzdy+ydw). \label{1_form_almost_Kahler}
\end{align}
It can be checked that $\cF$ is also co-closed (thus fulfilling~\eqref{F_co_closed}). Therefore, we can choose \eqref{example_base} with the almost-K\"{a}hler form \eqref{almost_Kahler_example} as the base space for the seed vacuum Taub-NUT metric and define $\mathbf{k}$ by \eqref{NUT_KS_vector} with the $1$-form \eqref{1_form_almost_Kahler}, thus generating a charged Taub-NUT solution with a single WAND $\mathbf{k}$ (a second single WAND $\mathbf{n}$ is given by~\eqref{coframe_charged_NUT}).

\subsubsection{Example with a double WAND $\mathbf{k}$ (Weyl type~II)}

The K\"{a}hler $2$-form for metric~\eqref{example_base} reads \cite{1999CQGra..16L...9N}
\begin{align}
   \cF =  \frac{l}{2}\left[dx\wedge(zdy -ydz+2dw)-2xdy\wedge dz\right] , \label{Kahler_example}
\end{align}
whose $1$-form $\mathbf{Z}$ can be chosen to be 
\begin{align}
    \mathbf{Z}=\frac{lx}{2}(zdy-ydz+ 2dw).\label{Kahler_1_form}
\end{align}
Using \eqref{Kahler_example} and \eqref{Kahler_1_form}, we have an example of charged Taub-NUT again with the same base space \eqref{example_base}, but now with $\mathbf{k}$ specified by \eqref{Kahler_1_form}, thus being a double WAND (a second double WAND $\mathbf{n}$ is given by~\eqref{coframe_charged_NUT}).

\section{Twistfree solutions}

\label{non_twisting_section}

When $\mathbf{k}$ is twistfree, from \eqref{F_compts} we see that $F_{ij} = 0$, and hence the CS current \eqref{CS_current} vanishes identically. As in the expanding-twisting case (Proposition~\ref{prop_charging_GKS}), this shows that also the twistfree case is immune to the CS term in the theory (but contrary to Proposition~\ref{prop_charging_GKS}, this is now true both for non-zero \cite{OrtPodZof08} and zero expansion, and odd dimensions can also occur). Two subclasses arise according to whether the shear vanishes.

\subsection{Shearfree solutions (Robinson-Trautman class)}

\label{RT_GKS_Subsection}

Here necessarily $m=n-2$ (cf.~Section~\ref{struc_opt_mat}). Since $\mathbf{k}$ in this case forms an expanding, twistfree, shearfree, null geodesic congruence (and thus satisfies the optical constraint \cite{OrtPraPra09,OrtPraPra13rev}), these solutions belong to the Robinson-Trautman class \cite{Robinson:1960zzb, Robinson:1962zz,Stephanibook,Griffiths_Podolský_2009}, extended to higher dimensions in \cite{Podolsky_RT}. Relevant to the present context is the work \cite{OrtPodZof08}, where the complete family of higher-dimensional Robinson-Trautman spacetimes solving the Einstein equations coupled to a $2$-form Maxwell(-CS) field $\mathbf{F}$ was obtained, under the assumption that $\mathbf{F}$ is aligned with the Robinson-Trautman null congruence (i.e., $F_{ab}k^b \propto k_a$). In the GKS case considered here, such an assumption is satisfied thanks to~\eqref{F_compts}. Therefore, to determine the complete class of higher-dimensional GKS solutions with an expanding, twistfree, shearfree, geodesic WAND $\mathbf{k}$, charged by an aligned $\mathbf{A}$ (i.e.~\eqref{aligned_A}), it is sufficient to identify the subclass of solutions of \cite{OrtPodZof08} for which $F_{ij} = 0$ (cf.~\eqref{F_compts}) and check whether the corresponding $\mathbf{F}$ can be generated by \eqref{aligned_A} and the metric belongs to the GKS class.\footnote{It may be useful to observe that, for a general geodesic null vector, the assumption of an aligned $\mathbf{F}$ (as in \cite{OrtPodZof08}) is weaker than assuming an aligned $\mathbf{A}$ (as in the present work). In particular, in the former case a non-vanishing magnetic part $F_{ij} \neq 0$ is permitted even in the twistfree case \cite{OrtPodZof08} (in which case the Einstein equations require $n$ to be even \cite{OrtPodZof08}). An example adding twist to such magnetic solutions of \cite{OrtPodZof08} will be presented in Appendix~\ref{subsec_app_NUT_nonalign}. \label{footn_A_aligned}}

Now, the solutions of \cite{OrtPodZof08} in the $F_{ij}=0$ branch are given by
\begin{align}
    &\mathbf{ g}=  r^2 \mathbf{h}(x) +2dudr -\left(K - \frac{2\Lambda}{(n-2)(n-1)}r^2 -\frac{\mu}{r^{n-3}} + \frac{n-3}{n-2}\frac{\kappa \alpha_0^2}{r^{2(n-3)}}\right)du^2, \label{2_form_charge_RT}\\
    &  \mathbf{F} = -\frac{(n-3)\alpha_0}{r^{n-2}}dr\wedge du,\label{F_RT}
\end{align}
where $K=0, \pm1$, $\mu$, $\alpha_0$ are constants, and the base space $\mathbf{h}=h_{ij}(x)dx^idx^j$ can be any Riemannian Einstein space with Ricci scalar equal to $K(n-2)(n-3)$ (in the KS subcase, the base space must additionally be of constant curvature \cite{OrtPraPra09,Ort_Srini}). The Robinson-Trautman congruence is defined by $\mathbf{k}=du$. It is then easily seen that the above solution belongs to the GKS class~\eqref{GKS}, where $\mathbf{\bar g}$ is given by~\eqref{2_form_charge_RT} with $\alpha_0=0$ (and is itself also GKS), and $2H=\frac{n-3}{n-2}\frac{\kappa \alpha_0^2}{r^{2(n-3)}}$. Furthermore, the field strength \eqref{F_RT} can be generated from the following vector potential
\begin{align}
    \mathbf{A}=\frac{\alpha_0}{r^{n-3}}\mathbf{k},
\end{align}
which is indeed consistent with the intermediate expression~\eqref{alpha_nontwist} (with $m=n-2$). Therefore, the solution~\eqref{2_form_charge_RT}, \eqref{F_RT} defines the complete class of charged GKS solutions with an expanding, twistfree and shearfree $\mathbf{k}$. These spacetimes are of Weyl type~D and $\mathbf{k}$ is one of the double WANDs \cite{OrtPodZof08}.\footnote{The additional branch of solutions of \cite{OrtPodZof08} with $F_{ij}\neq0$ mentioned in footnote~\ref{footn_A_aligned} is allowed only when the base space $\mathbf{h}$ is (almost-)K\"{a}hler, and hence only when $n$ is even. In that case, the field strength can be obtained from a (non-aligned) vector potential of the form $\mathbf{A} = \alpha_0r^{3-n}du - Z_i(x)dx^i$, where $\mathbf{Z}=Z_idx^i$ is a $1$-form associated with the (almost-)K\"{a}hler structure on $\mathbf{h}$.\label{footnote_aligned_F_RT}} The charged solutions~\eqref{2_form_charge_RT}, \eqref{F_RT} were first obtained in \cite{Tangherlini63,GibWil87}.

\subsection{Shearing solutions}

\label{subsec_shearing}

When $\mathbf{k}$ is expanding, twistfree and shearing, one must have $D \alpha=0$, the rank of the optical matrix must be $m=2$ (see Appendix \ref{shearing_appendix}), and $\mathbf{F}$ must be a {\em null} field (cf.~\eqref{F_compts}).\footnote{In the four-dimensional GKS solutions constructed in \cite{Kupeli88}, $\mathbf{k}$ is geodesic, twistfree, expanding and shearing, yet the electromagnetic field is non-null. However, this does not contradict our conclusion, since the GKS vector field $\mathbf{k}$ of \cite{Kupeli88} is not a WAND, and thus violates our assumption~\ref{ass4} (Section~\ref{subsec_backgr}).}
 This branch of solutions is therefore qualitatively different from the rest of the configurations considered in the present contribution, and we will thus not attempt a general study in arbitrary dimensions here. We limit ourselves to obtaining the full class for the specific case of five dimensions, additionally assuming $\mathbf{k}$ to be an mWAND -- the details are deferred to Appendix~\ref{shearing_appendix}. The analysis therein reveals that, necessarily, $\Lambda=0$. In what follows, we use the results of Appendix~\ref{shearing_appendix} to construct an explicit example.

\subsubsection{An example for $n=5$}

\label{subsibsec_5D_example}

Consider the following solution, where the metric is obtained from eqs.~\eqref{shear_twisfree_metric_final}, \eqref{U_bar_final} by setting $P=1$, with $C= c_1(u) x_1 + c_2(u) x_2$, and the field strength is obtained from eq.~\eqref{shearing_F} by choosing $\alpha = a_1(u) x_1 + a_2(u) x_2$ 
\begin{align}
   & \mathbf{ g} = 2\Big(\frac{ \mathcal {\bar M}(u)+M(u)}{r}\Big )du^2 -2dudr +  r^2 (dx_1^2+dx_2^2)   + \Big[dy + ( c_1 x_1 + c_2  x_2)du\Big ]^2, \label{example_5d}\\
   &  \mathbf{F}= -(a_1 dx_1 + a_2 dx_2)\wedge du , \label{example_5d_F} 
\end{align}
with $\mathbf{k}=-du$ (the choices of $C$ and $\alpha$ solve eqs.~\eqref{C_eq_final} and \eqref{alpha_shearing_PDE}, respectively.) The expressions for $\mathcal {\bar M}(u)$ and $M(u)$ are obtained respectively from \eqref{Mbar_lambda_0_1} (with $k=0$), \eqref{Ruu_background_lambda_0_1_2} and \eqref{M_lambda_0_soln_1_k_0} and read
\begin{align}
    \mathcal {\bar M}(u)=N(u)=-\frac{1}{4}\int du(c_1^2 +c_2^2)+\mu_0, \qquad
    M(u)=-\frac{\kappa}{2}\int du (a_1^2+a_2^2),
\end{align}
where $\mu_0$ is a constant and, without loss of generality, we have also absorbed the additive constant~$\eta$ from the $M(u)$ integral into $\mu_0$.

The (vacuum) background $\bar{\mathbf{g}}=\mathbf{g}|_{M=0}$ belongs to the class of solutions obtained by the Kaluza-Klein lift of the $n=4$ Robinson-Trautman solutions with a null electromagnetic field \cite{Reall:2012ih}, while the full geometry includes the backreaction of the null field~\eqref{example_5d_F}, amounting to the following pure radiation energy-momentum tensor (see~\eqref{shearing_EM_tensor})
\begin{align}
    & \mathbf{T}=\frac{1}{r^2}(a_1^2 +a_2^2)du^2 .
\end{align}

Since the background $\mathbf{\bar g}$ is generically not a KS metric, the full metric \eqref{example_5d} cannot be KS either \cite{Srinivasan:2025hro}, and therefore provides an example of a strictly GKS metric. The background metric $\mathbf{\bar g}$ is KS iff $c_1=0=c_2$ (and thus of type~D, or flat if, additionally, $\mu_0=0$), in which case $\mathbf{ g}$ reduces to the type~II shearing solution with a flat extra dimension discussed in \cite{Ort_Srini}.

\section*{Acknowledgments}
This work has been supported by the GA25-15544S grant from the Czech Science Foundation and the research plan RVO 67985840 of the Institute of Mathematics, Czech Academy of Sciences. A.S. is also grateful for additional support from the Charles University Research Center Grant No. UNCE24/SCI/016.

\renewcommand{\thesection}{\Alph{section}}
\setcounter{section}{0}

\renewcommand{\theequation}{{\thesection}\arabic{equation}}

\section{Ricci rotation coefficients and Ricci tensor (geodesic $\mathbf{k}$) }
\label{Curvature_appendix}

\setcounter{equation}{0}

Using the null frame for $\mathbf{g}$ and the GKS structure \eqref{GKS}, one can also define a null frame for the background $\mathbf{\bar g}$ as \cite{OrtPraPra09, Malek:2010mh,Ort_Srini,Srinivasan:2025hro}
\begin{align}
   \{\mathbf{\Bar{m}}_{(0)} = \mathbf{k}, \mathbf{\Bar{m}}_{(1)}= \mathbf{\Bar{n}}, \mathbf{\Bar{m}}_{(i)}=\mathbf{m}_{(i)}\},\label{frame_background}
\end{align}
where the null vector field $\bar n^a$ is defined as
\begin{align}
   \Bar{n}^a= n^a - H k^a, \label{n_bg}
\end{align}
and hence the corresponding covector field is given by $\Bar{n}_a= n_a + H k_a$. The following notation will be used to distinguish frame-projected quantities in the two frames:
\be
	T_{b_1 \ldots b_n} \equiv T_{a_1 \ldots a_n} m^{a_1}_{(b_1)} \ldots m^{a_n}_{(b_n)}, 
\qquad 
\bar{T}_{b_1 \ldots b_n} \equiv \bar{T}_{a_1 \ldots a_n} \bar{m}^{a_1}_{(b_1)} \ldots \bar{m}^{a_n}_{(b_n)} ,
\ee
where $T_{b_1 \ldots b_n}$ and $\bar{T}_{b_1 \ldots b_n}$ represent the projected components of the indexed object $T_{a_1 \ldots a_n}$ and its background counterpart $\bar{T}_{a_1 \ldots a_n}$, respectively.

Thus, one can also define the frame-projected covariant derivatives and the Ricci rotation coefficients for the background, analogous to equations \eqref{Frame_derivatives_GKS} and \eqref{GKS_Ricci_rot}. The Ricci rotation coefficients for $\mathbf{\bar g}$ and $\mathbf{g}$ are then related as \cite{Ort_Srini, Srinivasan:2025hro}
\begin{align}
  & L_{i0}=\Bar{L}_{i0} , \quad L_{10}=\Bar{L}_{10} , \quad L_{ij}=\Bar{L}_{ij} , \quad \overset{i}{M}_{j0}=\overset{i}{\Bar{M}}_{j0}, \quad \overset{i}{M}_{jk}=\overset{i}{\Bar{M}}_{jk},\label{Li0_GKS} \\
  & N_{i0}=\Bar N_{i0} , \quad L_{i1}=\Bar L_{i1} , \quad L_{1i}=\Bar L_{1i}-H \Bar L_{i0} , \quad N_{ij}=\Bar N_{ij}+H\Bar L_{ji} , \label{Nij_GKS} \\
  &\overset{i}{M}_{j1}=\overset{i}{\Bar M}_{j1}+H\big(\overset{i}{\Bar M}_{j0}+2\Bar L_{[ij]}\big) , \quad  L_{11}=\Bar L_{11}-H\Bar L_{10}-\Bar D H , \label{L11_GKS} \\
  & N_{i1}=\Bar N_{i1}+H \big(\Bar N_{i0}+2\Bar L_{1i}-H \Bar L_{i0}-\Bar L_{i1}\big)+ \Bar \delta_i H .  \label{Ni1_GKS}
\end{align}
Similarly, we have the following for the frame covariant derivatives \cite{Ort_Srini, Srinivasan:2025hro}
\begin{align}
Df &= \Bar{D}f , \quad \delta_i f = \Bar{\delta}_i f , \quad \Delta f = \Bar{\Delta}f + H \Bar{D}f, \label{Cov_der_bg_vs_GKS}
\end{align}
for any scalar function $f$. 

The above relations have already been used in \cite{Srinivasan:2025hro} to obtain expressions for the Riemann, Ricci, and Weyl tensor components in terms of the corresponding background curvature quantities. For our purposes, we recall below the ones for the Ricci tensor components, calculated assuming only a geodesic, affinely parametrized $\mathbf{k}$:\footnote{Beware of a sign error in the second term on the RHS of eqs. (B3), (B8), (B9), and (B12) of \cite{Srinivasan:2025hro}. In order to compare~\eqref{R11_GKS_1} with the corresponding expression~(B16,\cite{Ort_Srini}), one should recall~\eqref{Li0_GKS}--\eqref{L11_GKS}, \eqref{Cov_der_bg_vs_GKS} and use (11i,\cite{OrtPraPra07}).}
\begin{align}
      R_{00}=&\Bar R_{00} , \label{R00_GKS}\\
        R_{0i}=&   \Bar R_{0i}, \label{R0i_GKS}\\
       R_{ij} =&\Bar R_{ij} {+} 2H \Bar R_{0i0j} + 2HL_{ik}L_{jk} -2S_{ij}\left[DH+ (n-2)\theta H\right],\label{Rij_GKS}\\
     R_{01}  =& \Bar R_{01} {+} H \Bar R_{00} - [D^2H+ (n-2)\theta DH+2H\omega^2],\label{R01_GKS}\\
     R_{1i}  =& \Bar R_{1i} + H (\Bar R_{0i} + 2 \Bar R_{010i}) -\delta_i (DH)+ 2L_{[i1]} D H + 2L_{ij}\delta_{j}H - L_{jj}\delta_{i}H\nonumber\\
    &+ 2H(\delta_{j}A_{ij}+A_{ij}\overset{j}{M}_{kk}-A_{kj}\overset{i}{M}_{kj}- L_{jj}L_{1i} + 3L_{ij}L_{[1j]} + L_{ji}L_{(1j)}),\label{R1i_GKS}\\
    R_{11}= & \Bar R_{11}+H^{2}\Bar R_{00}   +\delta_{i}\delta_{i}H +(4L_{1i}-2L_{i1}+\overset{i}{M}_{kk})\delta_{i}H+\Bar N_{ii}DH-S_{ii}\Bar \Delta H \nonumber \\
 & +2H\Big(\delta_{i}L_{1i}- \Bar \Delta S_{ii}+4L_{1i}L_{[1i]}-L_{ki} \Bar N_{ki}+L_{1k}\overset{k}{M}_{ii}\Big).\label{R11_GKS_1}
\end{align}

The Ricci scalar, $R^a_{\hspace{1mm}a}$, reads
\begin{align}
    R=  \Bar{R}  {+} 4H \Bar R_{00} -2 \left[D^2H+ 2(n-2)\theta D H + H(n-2)(n-3)\theta^2 + H (\omega^2- \sigma^2)\right].
\end{align}

\section{Taub-NUT spacetimes in $n>4$: GKS structure and charging}
\label{appendix_NUT}
\setcounter{equation}{0}

In this appendix we review some aspects of the vacuum Taub-NUT metrics in higher dimensions \cite{Berard82,Bais:1984xb,PagPop87} (cf. also \cite{Lor_Dieter,ChaGib96,Taylor:1998fd,Awad:2000gg,Mann:Nuttier}), and provide a derivation of their charged generalization by means of a GKS transformation, as per Proposition \ref{prop_charging_GKS}. The analysis here will closely follow \cite{Ort_Srini}; however, unlike in \cite{Ort_Srini}, the derivation will not assume intersection with the KS class and will therefore be more general.

\subsection{Ansatz}

\label{subsec_NUT_ansatz}

To make its GKS structure manifest, the general Taub-NUT metric line-element of \cite{Bais:1984xb,PagPop87} can be written as \cite{Ort_Srini}
\be 
	d s^2=d r\otimes\mathbf{k}+\mathbf{k}\otimes d r+C^2(r)\mathbf{h} -2\H(r)\mathbf{k}\otimes\mathbf{k} , \qquad \mathbf{k}=d u-2\bZ , 
	\label{NUT1} 
\ee 
where $\mathbf{k}$ is a null covector field that will soon be identified with a WAND of the spacetime, $\mathbf{h}$ is the metric on the $(n-2)$-dimensional Riemannian base space with $x^\alpha$ denoting the corresponding coordinates, and $\mathbf{Z}=Z_\alpha (x) dx^\alpha$ is a 1-form on it. The functions $C$ and $\mathcal{H}$ are at this point arbitrary (they will be determined by the vacuum Einstein equations in Section~\ref{subsec_NUT_vac}).

Let us choose the null coframe to be
\be
 \mathbf m^{(0)}= dr-\mathcal{H}\mathbf{k}\equiv\mathbf{\bar n} , \qquad \mathbf m^{(1)}=\mathbf{k} , \qquad \mathbf m^{(i)}=C\mathbf{\tilde m}^{(i)},
 \label{coframe}
\ee
where $\mathbf{\tilde m}^{(i)}$ is an orthonormal coframe on the base space $\mathbf{h}$. It can then be shown that \cite{Ort_Srini}\footnote{The full list of non-vanishing Ricci rotation coefficients can be found in \cite{Ort_Srini}. Note, in particular, that the null coframe~\eqref{coframe} is not parallelly transported along $\mathbf{k}$, since $\overset{i}{M}_{j0}\neq 0$. While solving the Einstein-Maxwell equations in Section~\ref{charge_NUT_appnedix}, we will switch to a parallelly transported frame (cf. \cite{Ort_Srini}) to align with the main text -- but at this point the choice~\eqref{coframe} is more convenient.}
\beqn
 & & L_{ij}=C'C^{-1}\delta_{ij}+C^{-2}{\cal F}_{\tilde i\tilde j} , \qquad L_{i0}=0=L_{10}, 
 \label{opt_geod_NUT} 
\eeqn 
where $\cF \equiv \d\bZ$, the tilde over the spatial indices indicates components with respect to $\mathbf{\tilde m}^{(i)}$, and a prime denotes differentiation w.r.t $r$. Therefore, $\mathbf{k}$ is geodesic and affinely parametrized, and the coordinate $r$ can be identified with an affine parameter, since $k^a\pa_a=\pa_r$. Moreover, eq.~\eqref{opt_geod_NUT} reveals that $\mathbf{k}$ is expanding (provided $C'\neq0$), twisting, and {\em shearfree} ($\mathbf{\bar n}$ has the same properties \cite{Ort_Srini}; cf. also \cite{OrtPraPra13,Ortaggio17,Alekseevskyetal21,Taghavi-Chabert22}).

The non-zero frame components of the Riemann tensor are given by \cite{Ort_Srini} (cf. also \cite{PagPop87,Taghavi-Chabert22}) 
\beqn
  & & R_{0i0j}=C^{-4}\big(-C^3C''\delta_{ij}+{\cal F}_{\tilde i\tilde k}{\cal F}_{\tilde j\tilde k}\big) , \qquad R_{0ijk}= -2C^{-3}\tilde\nabla_{[\tilde k}{\cal F}_{\tilde j]\tilde i} , \label{R0i0j} \\
	& & R_{01ij}=-\big(2\H C^{-2})'{\cal F}_{\tilde i\tilde j} , \qquad R_{0101}=\H'' ,  \label{R01ij} \\
	& & R_{0i1j}=-\big(\H C^{-2})'{\cal F}_{\tilde i\tilde j}-C^{-1}\big(\H C'\big)'\delta_{ij}-C^{-4}\H{\cal F}_{\tilde i\tilde k}{\cal F}_{\tilde j\tilde k} , \label{R0i1j} \\ 
	& & R_{ijkl}=C^{-2}\tilde R_{\tilde i\tilde j\tilde k\tilde l}+4\H C ^{-4}\big[-\big(CC'\big)^2\delta_{i[k}\delta_{l]j}+{\cal F}_{\tilde i\tilde j}{\cal F}_{\tilde k\tilde l}-{\cal F}_{\tilde i[\tilde k}{\cal F}_{\tilde l]\tilde j}\big] , \label{Rijkl} \\	
	& & R_{1i1j}=\H^2R_{0i0j}  , \qquad R_{1ijk}=-\H R_{0ijk} . \label{R1i1j}
\eeqn

The non-trivial frame components of the Ricci tensor read \cite{Ort_Srini} (cf. also~\cite{Berard82,Bais:1984xb, PagPop87, Taghavi-Chabert22})
\beqn
  & & R_{00}=C^{-4}\big[-(n-2)C^3C''+{\cal F}^2\big] , \qquad R_{0i}= C^{-3}\tilde\nabla_{\tilde k}{\cal F}_{\tilde k\tilde i} ,\label{R00} \\	
	& & R_{01}=-\H''-(n-2)C^{-1}\big(\H C'\big)'-C^{-4}\H{\cal F}_{\tilde i\tilde k}{\cal F}_{\tilde i\tilde k} ,  \\
	& & R_{ij}=C^{-2}\big(\tilde R_{\tilde i\tilde j}+4\H C ^{-2}{\cal F}_{\tilde i\tilde k}{\cal F}_{\tilde j\tilde k}\big)-2\delta_{ij}C^{-2}\big[C(\H C')'+(n-3)\H(C')^2\big] , \\	
	& & R_{11}=\H^2R_{00}  , \qquad R_{1i}=-\H R_{0i} . \label{R11}
\eeqn

In the above expressions, $\tilde\nabla$ and $\tilde R_{\tilde i\tilde j}$ are, respectively, the covariant derivative and the Ricci tensor associated to (and projected along an o.n. frame of) $\mathbf{h}$, and $\mathcal{F}^2 \equiv \mathcal{F}_{\tilde i\tilde j} \mathcal{F}_{\tilde i\tilde j}$.

For later purposes, let us also display the following frame component of the Weyl tensor
\be
 (n-1)C_{0101} = (n-3) \Big[ \mathcal{H}'' - 2C^{-1}(\mathcal{H}C')' + 2C^{-2}(C')^2\mathcal{H} \Big] -\frac{2n}{n-2}C^{-4}\mathcal{H}{\cal F}^2 - \frac{\tilde R}{n-2}C^{-2} . 
 \label{C0101_NUT}
\ee

\subsection{Vacuum solutions}

\label{subsec_NUT_vac}

The $(00)$ and $(0i)$ components of the vacuum Einstein equations $R_{ab}=\frac{2\Lambda}{n-2}g_{ab}$ lead to \cite{Bais:1984xb,PagPop87,Ort_Srini}
\begin{align}
   & C^2= r^2+ l^2, \qquad l^2\equiv \frac{{\cal F}^2}{n-2}=\mbox{const} , \label{F^2_condition}\\
   &\tilde\nabla_{\tilde k}{\cal F}_{\tilde k\tilde i}=0 , \label{F_co_closed}
\end{align}
meaning that $\cF$ is co-closed. In addition, $R_{11}=0=R_{1i}$ are automatically satisfied thanks to~\eqref{R11}.

From now on, we shall assume that $\mathbf{k}$ defines a WAND of the geometry~\eqref{NUT1} (and so does then also $\mathbf{\bar n}$, cf.~\eqref{R1i1j}). By the first of~\eqref{R0i0j}, this means \cite{Ort_Srini} 
\be
 {\cal F}_{\tilde i\tilde k}{\cal F}_{\tilde j\tilde k}=l^2\delta_{ij} , 
 \label{WAND_condition_NUT}
\ee
which is equivalent to $\mathbf{k}$ obeying the {\em optical constraint} (see~\eqref{opt_geod_NUT} and cf. also \cite{OrtPraPra13rev}).\footnote{More generally, the Sachs equations \cite{OrtPraPra07} imply that, for a geodesic shearfree null direction, the WAND condition is equivalent to the optical constraint (not only for Taub-NUT spacetimes).} Together with the fact that $\cF$ is a closed 2-form, it also follows that the base space geometry $\mathbf{h}$ is (almost-)K\"{a}hler, and hence $n$ is necessarily even.\footnote{As shown in \cite{Ort_Srini}, vacua of the form~\eqref{NUT1} such that $\mathbf{k}$ and $\mathbf{\bar n}$ are {\em not} WANDs are possible only in the special branch $\mathcal{H}C^{-2} = \text{const}$ \cite{PagPop87}, for which eq.~\eqref{WAND_condition_NUT} can be violated (in which case $n$ is not necessarily even and the base space metric is not (almost-)K\"{a}hler-Einstein). We will not consider this possibility in the following, since we are ultimately interested in GKS spacetimes with the KS vector being a WAND (see \cite{Ort_Srini} for more details). For completeness, let us only note that the very special case $\mathcal{H}C^{-2} = \text{const}$ and with~\eqref{WAND_condition_NUT} still holding coincides with the subcase of metric~\eqref{GKS_NUT_vacuum}, \eqref{vacu_H_NUT} with $\mu_0=0$ and $(n-1)\tilde R=2n\Lambda l^2$ \cite{Taghavi-Chabert22,Ort_Srini} (for which $\mathbf{k}$ and $\mathbf{\bar n}$ are indeed WANDs).\label{footn_special_NUT}}

Finally, the $(ij)$ component of the field equations implies that the base space is Einstein, i.e.,
\be
 \tilde R_{\tilde i\tilde j}=\frac{\tilde R}{n-2}\delta_{ij} , \label{Rij_base}
\ee
along with \cite{Bais:1984xb,PagPop87,ChaGib96,Taylor:1998fd,ManSte06,multi_NUT_Mann,Alekseevskyetal21,Taghavi-Chabert22} (cf. also \cite{Ort_Srini})
\begin{align}
  -2\mathcal{H}=&\frac{r}{(r^2+l^2)^{\frac{n-2}{2}}}\left[\mu_0+ \frac{1}{n-2}\int^r d\bar r \Big[\Big(2\Lambda-\frac{\tilde R}{\bar r^2+l^2}\Big)\frac{(\bar r^2+l^2)^{\frac{n}{2}}}{\bar r^2}\Big] \right], 
	\label{vacu_H_NUT} 
\end{align}
where $\mu_0$ is an integration constant (note that $2\H(r^2+l^2)^{(n-2)/2}$ is a polynomial of degree $n$ in $r$ \cite{PagPop87,ChaGib96,Alekseevskyetal21}). The remaining Einstein equation corresponding to the $(01)$ component is automatically satisfied as a consequence of the Ricci identity \eqref{Ricci_3}, when combined with the solution to the $(ij)$ equation.

To summarize, $n>4$ vacuum Taub-NUT metrics for which the preferred null directions $\mathbf{k}$ and $\mathbf{\bar n}$ are WANDs exist only in even dimensions and are given by 
\begin{align}
   \mathbf{g}_{\small NUT}=d r\otimes\mathbf{k}+\mathbf{k}\otimes d r+(r^2+l^2)\mathbf{h} -2\H(r)\mathbf{k}\otimes\mathbf{k}, \qquad \mathbf{k}=d u-2\bZ ,
	 \label{GKS_NUT_vacuum}
\end{align}
where the base manifold $\mathbf{h}$ is (almost-)K\"{a}hler-Einstein, with fundamental 2-form $\cF\equiv \d\bZ$ with~\eqref{F^2_condition}, and $\mathcal{H}$ is given by \eqref{vacu_H_NUT}. Let us further recall that $\mathbf{k}$ and $ \mathbf{\bar n}$ are double WANDs iff $\tilde \nabla_i \mathcal{F}_{\tilde j \tilde k}=0$ \cite{Ort_Srini} (cf.~\eqref{R0i0j}, \eqref{R1i1j}), in which case the base space is K\"{a}hler. Note also that the GKS structure of the vacuum Taub-NUT metric~\eqref{GKS_NUT_vacuum} is manifest  (with $\mathbf{k}$ identified as the associated KS vector), the background being simply given by~\eqref{GKS_NUT_vacuum}, \eqref{vacu_H_NUT} with $\mu_0=0$.

We observe that $R_{010i}=0$ (cf. Section~\ref{subsec_NUT_ansatz}) with~\eqref{F_co_closed} (i.e., $R_{0i}=0$) imply $C_{010i}=0$, while \eqref{R0i1j} with \eqref{WAND_condition_NUT} and \eqref{Rij_base} result in $(n-2)C_{0(i|1|j)}=C_{0101}\delta_{ij}$. Using~(2.35,\cite{Durkeeetal10}) (contracted with the null rotation parameters $z_iz_j$) reveals that no WANDs other than $\mathbf{k}$ and $ \mathbf{\bar n}$ exist unless $C_{0101}$=0. However, (using~\eqref{C0101_NUT}) the latter restriction turns out to be compatible with the vacuum condition~\eqref{vacu_H_NUT} only when $\mu_0=0$ and $(n-1)\tilde R=2n\Lambda l^2$, i.e., in the special case $(n-1)(n-2)\mathcal{H}=-\Lambda(r^2+l^2)$ (cf. also a related comment in footnote~\ref{footn_special_NUT}). Except (possibly) for this fine-tuned case, the only WANDs of spacetime~\eqref{GKS_NUT_vacuum} are thus defined by $\mathbf{k}$ and $\mathbf{\bar n}$.

\subsection{Charging the Taub-NUT metrics}

\label{charge_NUT_appnedix}

We start by noting that the vacuum Taub-NUT metrics $\mathbf{g}_{\small NUT}$ \eqref{GKS_NUT_vacuum} satisfy all the assumptions of Proposition \ref{prop_charging_GKS}, i.e., they are Einstein spacetimes with an expanding, twisting geodesic WAND $\mathbf{k}$ that satisfies the optical constraint (cf. Sections~\ref{subsec_NUT_ansatz} and \ref{subsec_NUT_vac}). Hence, the result of the proposition applies to these metrics. Crucially, $\mathbf{k}$ is also shearfree, therefore the charging of these metrics with a potential of the form \eqref{aligned_A} is not forbidden by the proposition. Let us thus charge the vacua~\eqref{GKS_NUT_vacuum} by performing a GKS transformation
\begin{align}
    \mathbf{g}= \mathbf{g}_{\small NUT} -2H \mathbf{k}\otimes \mathbf{k},
		\label{charge_NUT_ansatz}
\end{align}
and taking the electromagnetic potential as in~\eqref{aligned_A}. The functions $\alpha$ and $H$ are to be fixed by solving the Maxwell equations \eqref{Max1}--\eqref{Max2} (with $j^a=0$, since $n$ is even) and the Einstein equations \eqref{01_eqn}--\eqref{11_eqn}.

It is useful to observe that the line-element~\eqref{charge_NUT_ansatz} is still of the general form~\eqref{NUT1}, up to redefining $\mathcal{H}\mapsto\mathcal{H}+H$, therefore the comments of Section~\ref{subsec_NUT_ansatz} apply also here. However, it is now more covenient to use a coframe parallelly transported along $\mathbf{k}$, given by (and related to the previous one~\eqref{coframe} by a spatial rotation) \cite{Ortaggio17,Ort_Srini}
\begin{align}
 & \mathbf m^{(0)}= dr-(\mathcal{H}+H)\mathbf{k}\equiv\mathbf{ n} , \qquad \mathbf m^{(1)}=\mathbf{k}=du- 2\mathbf{Z} , \qquad \mathbf m^{(i)}=(r^2+l^2)^{\frac{1}{2}}X_j^{\phantom{i}i}\mathbf{\tilde m}^{(j)},\label{coframe_charged_NUT}\\
 &  X_i^{\phantom{i}j}=l (r^2+l^2)^{-\frac{1}{2}}(\delta_i^j+rl^{-2}{\cal F}_{\tilde i}^{\phantom{\tilde i}\tilde j}) ,
\end{align}
where $\mathbf{\tilde m}^{(i)}$ is again an orthonormal coframe of $\mathbf{h}$. The dual frame reads
\be
  k^a\pa_a=\pa_r , \qquad n^a\pa_a=\pa_u+\left[\mathcal{H}+H\right]\mathbf{k} , \qquad \mathbf m_{(i)} = (r^2+l^2)^{-\frac{1}{2}}X_i^{\phantom{i}j}(  \mathbf {\tilde {m}_{(j)}}+2Z_{\tilde j}\pa_u) .
	\label{frame_KS_NUT}
\ee
Along with~\eqref{parallel_transp}, the Ricci rotation coefficients needed in solving the Einstein-Maxwell equations are \cite{Ort_Srini}
\beqn
 & & L_{ij}=(r^2+l^2)^{-1}\big(r\delta_{ij}+{\cal F}_{\tilde i\tilde j}\big) , \qquad L_{i1}=0=L_{1i} ,  \label{Lij_charged} \\
 & & N_{ij}=\left[\mathcal{H}+H\right]L_{ji} , \qquad \qquad  L_{11}=-D(\mathcal{H}+H) . \label{Nij_charged}
\eeqn 
From~\eqref{Lij_charged} one finds
\begin{align}
    & \theta=\frac{r}{r^2+l^2}, \qquad A_{ij}= \frac{1}{r^2+l^2 }{\cal F}_{\tilde i\tilde j} . \label{opt_decomp_NUT}
\end{align}
Using~\eqref{s_A}, it follows that all the twist functions $(a^0_{(2\mu)})^2$ take the same constant value 
\be
 (a^0_{(2)})^2=(a^0_{(4)})^2=\ldots=(a^0_{(n-2)})^2=l^2 ,
 \label{a_shearfree}
\ee 
and $m=2p=n-2$, in agreement with the shearfree property (cf. Section~\ref{struc_opt_mat}).

The general solution to the Maxwell equation \eqref{Max1} for an expanding $\mathbf{k}$ (obtained in \cite{Ort_Srini}) is listed in Section~\ref{subsection_Maxwell_eqn}. Thanks to~\eqref{a_shearfree}, from~\eqref{rescale_alpha} with \eqref{beta_even}, \eqref{A_alpha} one obtains \cite{Ort_Srini} (cf. also \cite{Awad06,ManSte06,DehKoh06})
\be
  \alpha =\frac{r}{(r^2+l^2)^{\frac{n-2}{2}}} \Bigg[\alpha_{0}+  \beta_{0}\sum_{\mu=0}^{\frac{n-2}{2}} \binom{\frac{n-2}{2}}{\mu} \frac{l^{2\mu}}{n-3-2\mu}r^{n-3-2\mu}\Bigg] ,
	\label{alpha_KS_NUT}	
\ee
where the $r$-independent integration functions $\alpha_0$ and $\beta_0$ are to be fixed by solving the remaining Maxwell equations \eqref{Max3} and \eqref{Max2}.

To that end, let us note that the Ricci identity \eqref{Ricci_4}, using~the second of \eqref{R0i0j}, \eqref{F_co_closed}, \eqref{Lij_charged}, and \eqref{opt_decomp_NUT}, leads to\footnote{To be precise, eqs.~\eqref{R0i0j} and \eqref{F_co_closed} were expressed in the frame~\eqref{coframe}, but they take the same form also in the frame~\eqref{coframe_charged_NUT}. It is to be noted that \eqref{11k_red_2}, which is an important ingredient in solving \eqref{Max3}, was derived in \cite{Ort_Srini} under the assumption that $\mathbf{k}$ is a double WAND (concretely, using $R_{0ijk}=0$, cf.~\eqref{R0i0j}, \eqref{F_co_closed}). However, as we have shown here, such an assumption can be weakened, namely, it is sufficient to use the co-closedness of $\mathcal{F}_{ij}$ (i.e., $R_{0jji}=0$). From this point onward, all subsequent steps in solving the Einstein-Maxwell equations, as carried out in \cite{Ort_Srini}, can be followed without any changes. For convenience, we will quickly sketch them out.}
\be
 \delta_{j}A_{ji}+\M{k}{j}{j}A_{ki} +\M{k}{i}{j}A_{jk}=0 .
 \label{11k_red_2}
\ee 

On using \eqref{11k_red_2} with~\eqref{comm_dD}, eq.~\eqref{Max3} reduces to 
\be
  \left[\delta_iD+2A_{ji}\delta_j +(n-4)\theta\delta_{i}\right]\alpha=0 .
	\label{Max3_KS_NUT}
\ee
Plugging~\eqref{alpha_KS_NUT} into \eqref{Max3_KS_NUT}, for $n > 4$ one finds (see \cite{Ort_Srini} for more details)
\be
 \delta_i\beta_{0}=0 , \qquad \delta_i\alpha_{0}=0 .
 \label{alpha_i}
\ee
Lastly, using~\eqref{frame_KS_NUT}, \eqref{Lij_charged}, \eqref{Nij_charged}, \eqref{Max1}, and \eqref{alpha_i}, one can reduce the Maxwell equation~\eqref{Max2} as \cite{Ort_Srini}
\be
  \alpha_{,ru}=0 .
\ee
Therefore, $\alpha_0$ and $\beta_0$ are constants; hence $\alpha$ depends only on the coordinate $r$.

Moving on to the Einstein equations, recall (cf. Section~\ref{subs_Einstein_eqns}) that they have reduced to~\eqref{1i_eqn}, \eqref{11_eqn}, \eqref{Eii} (eq.~\eqref{Eij_tracefree} is automatically satisfied as a consequence of the vacuum seed metric $\mathbf{g}_{\small NUT}$ being consistent with Proposition \ref{prop_charging_GKS}). Let us thus consider the remaining equations.

Integrating~\eqref{Eii}, one obtains \cite{Ort_Srini}
\begin{align}
 2H=-r\frac{\tilde \mu_0-\kappa Q(r)}{(r^2+l^2)^{\frac{n-2}{2}}} , \label{H_KS_NUT} 
\end{align}
with
\begin{align}
    D\tilde \mu_0=0 , \qquad DQ=-r^{-2}(r^2+l^2)^{\frac{n-4}{2}}\left[2\alpha^2 l^2 +\textstyle{\frac{1}{n-2}}(r^2+l^2)^2(D\alpha)^2\right] . \label{DF}
\end{align}
Finally, the remaining equations \eqref{1i_eqn} and \eqref{11_eqn} with~\eqref{R1i_GKS} and \eqref{R11_GKS_1} lead to (recall that the background satisfies $\Bar R_{1i}=0=\Bar R_{11}$) \cite{Ort_Srini}
\begin{align}
    \delta_i \tilde \mu_0=0, \qquad  \tilde \mu_{0,u}=0.
\end{align}
Therefore, $\tilde \mu_0$ is a constant, which can be absorbed into the background~$\mathbf{g}_{\small NUT}$ by redefining the parameter $\mu_0$ in~\eqref{GKS_NUT_vacuum}.

Summing up, the most general charged Taub-NUT metric that can be generated by a GKS transformation~\eqref{charge_NUT_ansatz} of \eqref{GKS_NUT_vacuum} with a vector potential of the form~\eqref{aligned_A} is determined by the metric function
\be
 2H= \frac{\kappa r Q(r)}{(r^2+l^2)^{\frac{n-2}{2}}} , 
 \label{H_Q}
\ee
and the Maxwell field strength 
\be
 \mathbf{F}=(D\alpha) d r \wedge\mathbf k-2\alpha\cF , 
 \label{shearfree_F}
\ee
where $\alpha$ and $Q$ are functions of $r$ defined by \eqref{alpha_KS_NUT} and \eqref{DF}, respectively.

An argument similar to the one used in Section~\ref{subsec_NUT_vac} for the vacuum case reveals that $\mathbf{k}$ and $\mathbf{n}$ are the only WANDs (we observe that, for a non-zero Maxwell field, no possible fine-tuned exceptions can occur; we refer to \cite{Srinivasan_PhD} for more details).

\subsection{A six-dimensional example of charged Taub-NUT with a non-aligned vector potential}

\label{subsec_app_NUT_nonalign}

The alignment condition~\eqref{aligned_A} has been one of the key assumptions in the GKS charging method proposed in the present paper. However, the classical family of electrovac KS solutions of \cite{DebKerSch69} includes also potentials of a more general form (cf. also \cite{Stephanibook}). Similarly, certain higher dimensional solutions for which the metric is (G)KS but the potential is more general than~\eqref{aligned_A} are also known \cite{Ortaggio:RT_pform} (see footnotes~\ref{footn_A_aligned} and \ref{footnote_aligned_F_RT}, and \cite{Ort_Srini} for further comments). It is the purpose of the present section to show by means of an example that, at least in some cases, the GKS construction considered above can be extended in a similar direction. A more systematic analysis of such a possibility is left for future work.

Let us construct an example of charged Taub-NUT in $n=6$ of the GKS form~\eqref{GKS} but with a potential~$\mathbf{A}$ which does not adhere to the alignment assumption \eqref{aligned_A}. The ansatz for the vacuum seed is $\mathbf{\Bar{g}}=\mathbf{g}_{\small NUT}$ (given in~\eqref{GKS_NUT_vacuum_main}), with the base space being a flat Riemannian metric
\begin{align}
    \mathbf{h}= dx^2 +dy^2+dz^2+dw^2.
\end{align}

We can choose the K\"{a}hler $1$-form to be
\begin{align}
    \mathbf{Z}=\frac{l}{2}(xdy-ydx+zdw-wdz),
\end{align}
which then also defines the KS vector $\mathbf{k}$ through \eqref{NUT_KS_vector}. The vacuum function $\mathcal{H}$ is given by~\eqref{Hvac_6D} (the same as for the six-dimensional examples of Section~\ref{6d_examples_aligned_A}, since the base space considered there was also Ricci-flat). Starting from an educated ansatz and solving the Einstein--Maxwell equations, we have arrived at the following solution (after conveniently choosing the gauge for $\mathbf{A}$ and using the freedom $u\mapsto u+f(x,y)$)
\begin{align}
   2H=& \frac{\kappa}{(r^2+l^2)^{2}}\left[rQ(r)-\frac{q^2}{2}(r^2-l^2)\right]  ,  \\
	 \mathbf{A}=&\alpha(r)\mathbf{k}+ \frac{q}{2}(zdx-xdz+ydw-wdy) ,\label{non_aligned_A}
\end{align}
where $q$ is a constant and $\alpha(r)$, $Q(r)$ are as given by \eqref{alpha_soln_6d}, \eqref{Q_soln_6d}. The obtained $\mathbf{A}$ is not proportional to $\mathbf{k}$ since it contains an additional term proportional to a K\"{a}hler $1$-form different from $\mathbf{Z}$ -- in other words, the field strength $\mathbf{F}=d\mathbf{A}$ is not of the form \eqref{F_soln_NUT}. However, $\mathbf{F}$ itself is aligned with $\mathbf{k}$, in the sense that $k^aF_{a[b}k_{c]}=0$. For $\Lambda=0$, this becomes a KS solution extending an example given in Section~III.B.3 of \cite{Ort_Srini}.

Note that $\mathbf{F}$ retains its magnetic components $F_{ij}$ proportional to $q$ even in the twistfree limit $l \rightarrow 0$, which gives rise to one of the Robinson-Trautman solutions of \cite{OrtPodZof08}.

\section{Expanding-twistfree-shearing solutions}

\label{shearing_appendix}

\setcounter{equation}{0}

In this appendix we provide certain technical details relevant to Section~\ref{subsec_shearing}, i.e., to the case when the WAND $\mathbf{k}$ is expanding, twistfree (i.e., $p=0$ in~\eqref{L_ij_canonical}) and {\em shearing}. We will first obtain a few results in the general case, and subsequently the full family of solutions in $n=5$ dimensions when $\mathbf{k}$ is assumed to be an mWAND (thus not just a WAND) -- thus also extending the particular five-dimensional KS example presented in \cite{Ort_Srini}.

\subsection{Preliminary results}

Imposing the twistfree condition on the Einstein equation component~\eqref{Eij_tracefree}, thanks to $\theta\neq0$ one gets
\begin{align}
   D\alpha=0 ,
	 \label{twistfree_Eij_tracefree}
\end{align}
i.e., $\alpha$ must be $r$-independent, and thus $F_{01}=0$ (eq.~\eqref{F_compts}). As already noted in Section \ref{non_twisting_section}, $F_{ij}=0$ by virtue of vanishing twist. Therefore, $F_{1i}$ are the only possible nontrivial field strength components, meaning that $\mathbf{F}$ is a null field.

Recalling that the CS current is identically zero for the twistfree case (see Section \ref{non_twisting_section}), consider the Maxwell equation \eqref{Max3}, which upon using $D\alpha=0=A_{ij}$, eqs.~\eqref{comm_dD}, \eqref{F_compts}, and $(n-2)\theta=mr^{-1}$ (cf.~\eqref{exp}), can be rewritten as
\begin{align}
 \left(2L_{ij}-\frac{m}{r}\delta_{ij}\right)F_{1j}=0 . 
 \label{F_1i_shearing_Maxwell}
\end{align}

The canonical form of the optical matrix \eqref{L_ij_canonical}--\eqref{diag_block_Lij_canonical} for the expanding-twistfree-shearing case reads 
  \begin{align*}
     L_{ij}=\frac{1}{r}\mbox{diag}(\underbrace{1,\ldots,1}_{m},\underbrace{0,\ldots,0}_{(n-2-m)}) 
,  
  \end{align*}
with $1\le m\le n-3$. Therefore, eq.~\eqref{F_1i_shearing_Maxwell} takes the following forms for the respective values of $i$
\begin{align}
       &\frac{m-2}{r }F_{1i}=0, \qquad i=2, \dots, m+1 , \\
       &\frac{m}{r }F_{1i}=0, \qquad i=m+2,\dots, n-1 .
\end{align}
In order for the field strength to be non-trivial one thus finds
\be
 m=2 ,
\ee
and the only possible non-zero components in the above canonical frame are $F_{12}$ and $F_{13}$.

Next, the Einstein equation~\eqref{Eii} (using also~\eqref{shear}) reduces to $DH= -\frac{H}{r}$, and thus
\begin{align}
	H=-\frac{M}{r} ,
	\label{M_shearing}
\end{align}
where $M$ is an $r$-independent integration function, to be determined using the remaining Einstein equations. The Einstein equation~\eqref{Eij_tracefree} is identically satisfied, so that one needs to solve only the components $R_{1i}$  (with $T_{1i}=0$, cf.~\eqref{1i_eqn} and (47,\cite{Ort_Srini})) and $R_{11}$ (with (48,\cite{Ort_Srini})).

\subsection{The full family of $n=5$ solutions with $\mathbf{k}$ forming an mWAND}

Our starting point is the following background seed for the GKS transformation, which forms the most general five-dimensional metric characterized by a geodesic, expanding, twistfree, shearing mWAND $\mathbf{k}=-du$ with a rank-2 optical matrix, cf.~(55,\cite{Reall:2012ih}) 
\begin{align}
\mathbf{\bar g} = -2\bar U du^2 -2dudr + r^2P^{-2}\delta_{\alpha \beta }(d x^\alpha + B^\alpha du)(d x^\beta + B^\beta du) + \big[dy + \big(C+ 2r(\ln P)_{,y}\big)du\big ]^2,  \label{shear_twisfree_metric_5d}
\end{align}
where $\alpha, \beta, \dots =1,2 $. 
The metric functions $\bar U(u, r, x^\alpha, y)$, $B^\beta(u, x^\alpha, y)$, $C(u, x^\alpha, y)$, and $P(u, x^\alpha, y)>0$ are constrained by the vacuum Einstein equations \cite{Reall:2012ih}, while the charged metric $\mathbf{g}$, resulting from the GKS transformation, preserves the form \eqref{shear_twisfree_metric_5d} with $\bar U \rightarrow U = \bar U + H$ and \eqref{M_shearing}. The optical matrix of $\mathbf{k}$ obeys the optical constraint \cite{Ortaggioetal12}, in agreement with assumption~\ref{ass3}. in section~\ref{subsec_backgr}.

\subsubsection{Maxwell equations}

Owing to~\eqref{twistfree_Eij_tracefree}, the aligned vector potential \eqref{aligned_A} gives the following coordinate form of the field strength
\begin{align}
\mathbf{F}= -(\alpha_{,y}  dy + \alpha_{,\beta}  dx^\beta) \wedge du .  \label{shearing_F}
\end{align}
The Maxwell equations $\nabla_a F^{ab}=0$ thus reduce to
\begin{align}
	\alpha_{,y}=0 , \qquad (\partial_{x^1}^2 + \partial_{x^2}^2)\alpha=0. 
\label{alpha_shearing_PDE}
\end{align}

\subsubsection{Einstein equations}

From the form of the field strength, it is easy to see that the only non-trivial energy-momentum tensor component is given by
\begin{align}
T_{uu}=\frac{P^2}{r^2}(\nabla \alpha)^2 , \label{shearing_EM_tensor}
\end{align}
where $(\nabla \alpha)^2=(\alpha_{,x^1})^2 + (\alpha_{,x^2})^2$.
Hence, except for the $R_{uu}$ equation, all other Einstein equations remain unchanged from the vacuum case. For these, we can directly invoke the results of \cite{Reall:2012ih} (cf., in particular, eqs. (63)--(68) therein, noting a different normalization of $U$), namely 
\begin{align}
    & B^\alpha=0, \qquad  C=C(u, x^\alpha), \label{B_C_5d_shearing}\\
    & P_{,yy}+ \frac{\lambda}{4}P=0, \label{P_condition_1}\\
    & U=-\frac{\mathcal{M}}{r}+ \frac{1}{2}\Delta\ln P -r[(\ln P)_{,u}-C(\ln P)_{,y}]+ \frac{r^2}{2} \left(3[(\ln P)_{,y}]^2 -\frac{\lambda}{4} \right), \label{U_soln_1_5d}\\
     & \Delta C=-2\mathcal{ M}_{,y}+ 8\mathcal{ M} (\ln P)_{,y},  \label{m_eqn_1}\\
     & 12 \mathcal{ M} [(\ln P)_{,y}]^2 -6 \mathcal{ M}_{,y}(\ln P)_{,y} + \mathcal{ M}_{,yy} +\lambda \mathcal{ M}=0, \label{m_eqn_2}
\end{align}
where $\lambda\equiv\frac{2\Lambda}{n-2} = \frac{2\Lambda}{3}$, $\mathcal{M}=\mathcal{ M}(u,y)$ is an integration function, and $\Delta\equiv P^2 \delta^{\alpha \beta }\partial_\alpha \partial_\beta$.

The GKS transformation affects neither the background metric nor the cosmological constant. Therefore, imposing the GKS structure we can split $\mathcal{ M}$ as
\begin{align}
 M\equiv \mathcal{M}-\mathcal{\bar M},\label{GKS_fn_shearing}
\end{align}
where $\mathcal{\bar M}(u,y)$ is completely decoupled from the matter field so that 
\begin{align}
\bar U\equiv U|_{\mathcal{M}\mapsto\mathcal{\bar M}}
\end{align}
solves the vacuum equations, while the GKS function $H=U-\bar U$ coupled to matter is given by \eqref{M_shearing}. 
Comparing eqs.~\eqref{m_eqn_1}, \eqref{m_eqn_2} with their respective vacuum background counterparts \cite{Reall:2012ih} (obtained with $\mathcal{M} \mapsto \mathcal{\bar M}$), one finds that the GKS splitting \eqref{GKS_fn_shearing} imposes the following constraints on $M(u,y)$
\begin{align}
    &(\ln M)_{,y}=4(\ln P)_{,y}, \label{M_eqn_1}\\
    & 12 M [(\ln  P)_{,y}]^2 -6 M_{,y}(\ln  P)_{,y} + M_{,yy} +\lambda M=0 . \label{M_eqn_2}
\end{align}

Note that in equation \eqref{M_eqn_1}, the LHS is independent of $x^\alpha$. Therefore, $P$ must be of the following form
\begin{align}
    P= P_1(u,y) P_2(u, x^\alpha).\label{P_expression}
\end{align}
Solving \eqref{M_eqn_1}, we get
\begin{align}
    M(u,y)= P_1^4 g(u),\label{M_soln_1}
\end{align}
where $g(u)$ is a (non-zero) free function.

Let us now consider the $R_{uu}$ equation. First, for the background it is given by (69,\cite{Reall:2012ih}), i.e.,
\begin{align}
    \Delta \Delta\ln  P + 12\mathcal{\bar M}[(\ln  P)_{,u}-C(\ln  P)_{,y}]- 4(\mathcal{\bar M}_{,u}-C\mathcal{\bar M}_{,y})- P^2 (\nabla C)^2=0 .\label{Ruu_background}
\end{align}
Next, for the full spacetime with~\eqref{shearing_EM_tensor}, after some simplifications using~\eqref{P_condition_1}, \eqref{M_eqn_2}, and \eqref{M_eqn_1}, the condition resulting from the $R_{uu}$ equation reads
\begin{align}
   2MC(\ln  P)_{,y} + 6M(\ln  P)_{,u}-2 M_{,u}=\kappa P^2(\nabla \alpha)^2 . \label{R_uu_PDE_2}
\end{align}
Thanks to~\eqref{P_expression} and \eqref{M_soln_1}, eq.~\eqref{R_uu_PDE_2} can be rewritten as
\begin{align}
   2\bigg[CP_1P_{1,y} + P_1^2\left(\ln\frac{P_2^3}{g}\right)_{,u}- P_1P_{1,u}\bigg]=g^{-1}\kappa  P_2^2(\nabla \alpha)^2 . 
	 \label{M0}
\end{align}
Taking the $y$-derivative of \eqref{M0} and using \eqref{P_condition_1}, \eqref{P_expression}, we get
\begin{equation}
C\!\left[(P_{1,y})^2 - \frac{\lambda}{4}P_1^2\right]
+ 2P_1P_{1,y}\!\left(\ln\frac{P_2^3}{g}\right)_{,u}
- P_{1,u}P_{1,y}
- P_1 P_{1,uy}=0 .
\label{M1}
\end{equation}
Differentiating equation \eqref{M1} with respect to $y$ and using again equations \eqref{P_condition_1} and \eqref{P_expression} gives
\begin{equation}
-\lambda C\, P_1 P_{1,y}
+ 
2\left(\ln\frac{P_2^3}{g}\right)_{,u}\left[(P_{1,y})^2 - \frac{\lambda}{4}P_1^2\right]
+ \frac{\lambda}{2}P_1 P_{1,u}- 2P_{1,y}P_{1,yu}
=0 
\label{M2}
\end{equation}

A suitable linear combination of \eqref{M1} and \eqref{M2} yields 
\begin{align}
    \bigg[(P_{1,y})^2+\frac{\lambda}{4}P_1^2\bigg]\bigg[   CP_1P_{1,y} + P_1^2\left(\ln\frac{P_2^3}{g}\right)_{,u}- P_1P_{1,u}\bigg]=0 .
\end{align}
Since the second factor equals the bracketed expression on the LHS of~\eqref{M0}, one concludes that
\be
	(P_{1,y})^2+\frac{\lambda}{4}P_1^2=0 ,
\ee
or else $\alpha=\alpha(u)$ and the vector $\mathbf{A}=\alpha du$ would be a pure gauge.
This singles out two cases: 
		\begin{enumerate} 
			\item $\lambda=0=P_{1,y}$, or
			\item $\lambda<0$, $P_{1}=a(u)e^{\pm\frac{\sqrt{-\lambda}}{2}y}$. However, this means that the $y$-dependence factors out in the LHS of~\eqref{M0}, implying $(\alpha_{,x^1})^2 + (\alpha_{,x^2})^2=0$. Hence, we have $\alpha=\alpha(u)$, and thus the vector $\mathbf{A}=\alpha du$ is a pure gauge, ruling out this case.
		\end{enumerate} 
	Therefore, let us proceed to study the only remaining case, i.e., $\lambda=0=P_{1,y}$.

Since $P_1=P_1(u)$, without loss of generality we can set $P_1=1$ by absorbing the $u$-dependence into $P_2$, hence $P=P_2(u, x^\alpha)$ (from now on we will thus drop the subscript $_2$) and $M=M(u)(=g(u))$. Eq.~\eqref{R_uu_PDE_2} thus reduces to
\begin{align}
   \left(MP^{-3}\right)_{,u}=-\frac{\kappa}{2}P^{-1}(\nabla \alpha)^2 ,
	\label{R_uu_PDE_2_bis}
\end{align}
while the $\mathcal{\bar M}$ analogue of equation \eqref{m_eqn_1} becomes
    \begin{align}
          \Delta C=-2 \mathcal{\bar M}_{,y}.\label{m_bar_eqn_lambda_0_1}
    \end{align}
    The LHS is independent of $y$ and the RHS is independent of $x^\alpha$. Therefore, they must be equal to a function that depends only on $u$. Hence, we have
    \begin{align}
       \mathcal{\bar M}= k(u) y + N(u) , \label{Mbar_lambda_0_1}
    \end{align}
so that \eqref{m_bar_eqn_lambda_0_1} simplifies to
 	    \begin{align}
           \Delta C=-2k(u) .
					\label{C_eq}
    \end{align}
Using $P_{,y}=0$ and~\eqref{Mbar_lambda_0_1}, eq.~\eqref{Ruu_background} becomes
    \begin{align}
    \Delta \Delta\ln  P + 12[k(u) y + N(u)](\ln  P)_{,u}- 4(yk_{,u}+N_{,u}-Ck)- P^2 (\nabla C)^2=0.\label{Ruu_background_lambda_0_1}
\end{align}
The $y$-dependent terms in \eqref{Ruu_background_lambda_0_1} must sum to zero, i.e.,
\begin{align}
    3 k (\ln  P)_{,u}- k_{,u}=0.
		\label{k_u}
\end{align}
This splits into two subcases: (i) $k\neq 0 $ and (ii) $k=0$. In the following, we first show that the former leads in fact to an inconsistency, and we then work out the latter.

\paragraph{(i) $k\neq 0$:} As in \cite{Reall:2012ih}, a suitable rescaling of $u$ and $r$, and a shift of $y$ can be used to set 
\be
 k=\mbox{const}\neq0, \qquad N(u)=0 ,
\ee
i.e., $P_{,u}=0$ (cf.~\eqref{k_u}) and $\mathcal{\bar M}= ky$, which will thus be assumed hereafter.

 Eq.~\eqref{Ruu_background_lambda_0_1} now reads
 \begin{align}
    \Delta \Delta\ln P + 4Ck=P^2 (\nabla C)^2 , \label{Ruu_background_lambda_0_1_1}
\end{align}
while \eqref{R_uu_PDE_2_bis} becomes
\begin{align}
 \left(M\right)_{,u} = -\frac{\kappa}{ 2}P^{2}(\nabla \alpha)^2 . \label{Mu_pde_lambda_0_1} 
\end{align}
Note that the LHS depends only on $u$ and hence so must the RHS. We can thus label
\begin{align}
  \omega^2(u)\equiv \frac{1}{4}P^2(\nabla \alpha)^2 , 
	\label{h}
\end{align}
which must be non-zero if a non-vanishing Maxwell field is assumed.

Let us now switch to the complex coordinates $(z,\bar z)$ defined by
\begin{align}
    z= x_1+ i x_2 ,
\end{align}
such that $\Delta=4P^2\partial_z \partial_{\bar z}$. Eq.~\eqref{alpha_shearing_PDE} is thus solved by 
\begin{align}
   \alpha= f(u,z) +\bar f(u,\bar z) ,
\end{align}
where $f$ is holomorphic in $z$, while~\eqref{h} becomes
\begin{align}
      P^2f_{,z} \bar f_{,\bar z}=\omega^2 (u) .
			\label{P_f_constr}
\end{align}
Taking the logarithm of the latter and differentiating w.r.t.~$z$ and $\bar z$, one gets $\Delta \ln P=0$, such that~\eqref{Ruu_background_lambda_0_1_1} becomes (using~\eqref{P_f_constr})
\begin{align}
     C_{,z}C_{,\bar z} =   \frac{Ck}{\omega^2}f_{,z} \bar f_{,\bar z} . 
		\label{C_k_eqn_1}
\end{align}

From \eqref{C_eq} with \eqref{P_f_constr}, we have
\begin{align}
   C_{,z\bar z}=-\frac{k}{ 2\omega^2}f_{,z} \bar f_{,\bar z} , 
	\label{C_k_eqn_2}
\end{align}
which is solved by
\begin{align}
    C=-\frac{k}{ 2\omega^2}[f\bar f+h(u,z)+ \bar h(u,\bar z)].  \label{C_soln}
\end{align}
Substituting~\eqref{C_soln} into \eqref{C_k_eqn_1} and dividing both sides by $f_{,z}\bar f_{,\bar z}$ results in
\begin{align}
    3f\bar f +\frac{f}{f_{,z}}h_{,z}+\frac{\bar f}{\bar f_{,\bar z}}\bar h_{,\bar z}+ \frac{h_{,z}}{f_{,z}}\frac{\bar h_{,\bar z}}{\bar f_{,\bar z}}=-2(h+\bar h) .
		\label{f_H_eqn_2}
\end{align}
Differentiating w.r.t. $z$ and $\bar z$ gives
\begin{align}
    3f_{,z}\bar f_{\bar z}+ \left(\frac{h_{,z}}{f_{,z}}\right)_{,z}\left(\frac{\bar h_{,\bar z}}{\bar f_{,\bar z}}\right)_{,\bar z}=0 .
\end{align}
This is a sum of non-negative quantities, which must thus vanish separately, implying in particular $f_{,z}=0$. Hence, we arrive at a contradiction, and this case is ruled out.

\paragraph{(ii) $k=0$:} eq.~\eqref{Ruu_background_lambda_0_1} reduces to 
  \begin{align}
   \Delta \Delta\ln  P + 12N(u)(\ln  P)_{,u}- 4N_{,u}- P^2 (\nabla C)^2=0.\label{Ruu_background_lambda_0_1_2}
\end{align}
Eq.~\eqref{R_uu_PDE_2_bis} is solved as 
\begin{align}
    M(u)= P^3\left[-\frac{\kappa }{2} \int P^{-1}(\nabla \alpha)^2du   +\eta(x^\alpha)\right] ,\label{M_lambda_0_soln_1_k_0}
\end{align}
where the $x^\alpha$ dependence of $P$, $\eta$ and $\alpha$ must be such that the RHS is a function of only $u$.

\subsubsection{Summary}

To summarize, five-dimensional charged GKS solutions of~\eqref{action} with an expanding, twistfree, shearing, geodesic mWAND $\mathbf{k}$ are possible only for $\Lambda=0$. The full class is given by the metric
\begin{align}
&\mathbf{ g} = -2 \bar U du^2 -2dudr + \frac{r^2}{P^2}{(dx_1^2+dx_2^2)} +(dy + Cdu)^2 +\frac{2{M(u)}}{r}du^2,  \label{shear_twisfree_metric_final}
\end{align}
with $P=P(u,x^\alpha)$, {$C=C(u,x^\alpha)$}, and 
\beqn
& & \Delta C=0 , \label{C_eq_final} \\
& & \bar U=-\frac{N(u)}{r}+ \frac{1}{2} \Delta\ln  P -r(\ln  P)_{,u} , \label{U_bar_final}
\eeqn
so that $C$ is harmonic (in the special case $C=C(u)$, it can be set to zero by a shift of $y$, leading to a direct product metric). The Maxwell potential reads $\mathbf{A}=-\alpha du$, where $\alpha(u, x^\beta)$ satisfies \eqref{alpha_shearing_PDE}, i.e., it is also harmonic. The vector field $\pa_y$ is manifestly Killing.  The functions $P$ and $C$ are constrained by~\eqref{Ruu_background_lambda_0_1_2} and $M$ is given by \eqref{M_lambda_0_soln_1_k_0}. An explicit example in this branch is provided in Section~\ref{subsibsec_5D_example}. See \cite{Reall:2012ih} for more comments on the properties of the background spacetime.

To conclude, it is worth noticing that the above analysis reveals that, as a consequence of imposing the GKS structure, {\em not} all five-dimensional vacua of \cite{Reall:2012ih} admit a charged extension of the kind considered in the present paper. Recall, in particular, the additional constraints on the background functions given by~\eqref{P_expression} (with $P_1=1$), \eqref{Mbar_lambda_0_1} (with $k=0$), and the comment following~\eqref{M_lambda_0_soln_1_k_0}.

\section{Non-expanding $\mathbf{k}$ (Kundt class)}

\setcounter{equation}{0}

\label{app_Kundt}

Whereas the rest of the paper assumes $\theta\neq0$, this appendix is devoted to the remaining branch of solutions defined by $\theta=0$. This case was not considered in \cite{Ort_Srini}, therefore the results of this appendix provide new insight also for the KS case. As noted in Section~\ref{subsubsec_nonexp} the optical constraint~\eqref{OC} implies that $\mathbf{k}$ is Kundt, i.e., $L_{ij}=0$, and the resulting spacetime is GKS-Kundt, where the KS vector also defines a Kundt vector.

Thanks to $\theta=0=\omega$, the general solution to \eqref{Max1} here takes the form
\begin{align}
    \alpha= \alpha_0 r + \beta_0,  
		\label{alpha_Kundt_case}
\end{align}
where $\alpha_0$ and $\beta_0$ are $r$-independent integration functions, such that
\begin{align}
    \mathbf{F}
= \alpha_0 \mathbf{n}\wedge\mathbf{k} + F_{1i}\mathbf{k}\wedge \mathbf m^{(i)} ,
\label{F_Kundt_gen_frame}
\end{align}
with~\eqref{F_compts} and \eqref{alpha_Kundt_case}.

In the main text, the principal focus of this paper has been the analysis of solutions to~\eqref{action} of the form~\eqref{GKS}, where $\mathbf{\bar g}$ is also a solution of~\eqref{action} but with $\mathbf{F}=0$ (in addition to the remaining assumptions described in Section~\ref{sec_intro}). This means that $\Lambda$ represents the cosmological constant of both the background and the full spacetime and, to emphasize this, we shall write $\bar\Lambda=\Lambda$. Let us thus first consider this case also here. As a small detour from the main theme of the paper, the case when the cosmological constant of the background is allowed to differ from the cosmological constant of the full spacetime, i.e., $\bar\Lambda\neq\Lambda$ will be discussed in Section~\ref{subsec_app_Kundt_nonnull}.

\subsection{Case $\bar\Lambda=\Lambda$: null fields}

\label{subsec_app_Kundt_null}

The Einstein equation~\eqref{ij_eqn} with \eqref{Rij_GKS} and \eqref{alpha_Kundt_case} thus gives 
\be
 \alpha_0=0 ,
 \label{alpha_null}
\ee
which means that the Maxwell field~\eqref{F_Kundt_gen_frame} reduces to $\mathbf{F}=F_{1i}\mathbf{k}\wedge \mathbf m^{(i)}$ and is null.

Next, the Einstein equation~\eqref{01_eqn} (with~\eqref{R01_GKS}) requires
\begin{align}
    H= -\frac{1}{2}(b_H r + c_H), 
		\label{r_depend_H_Kundt}
\end{align}
where $b_H$ and $c_H$ are $r$-independent integration functions (the overall normalization factor $-\frac{1}{2}$ has been chosen to match the conventions of \cite{Podolsky:2008ec} in what follows). 

The rest of the analysis can be simplified considerably by noticing that general $n>4$ Kundt spacetimes in the Einstein-Maxwell theory with an aligned $\mathbf{F}$ have been already studied in \cite{Podolsky:2008ec}. It is also shown there that the Kundt vector forms an mWAND (which is a consequence of the alignment condition \cite{OrtPraPra07}). Therefore, the full family of solutions for the current case will be given by the subclass of those of \cite{Podolsky:2008ec} for which the field strength can be derived from an aligned $\mathbf{A}$~\eqref{aligned_A}, with the charging corresponding to a GKS transformation (after identifying $\mathbf{k}$ with the Kundt vector of \cite{Podolsky:2008ec}).

Let us introduce canonical Kundt coordinates $(u,r,x)$ (where $x$ collectively denotes the dependence on the transverse spatial coordinates $x^\a$, $\a,\beta=2,\ldots,n-1$), such that $\mathbf{k}=-du$ and $k^a\pa_a=\pa_r$ \cite{ColHerPel06,Coleyetal03,Podolsky_RT,Podolsky:2008ec,Podolsky:2013qwa} (cf.~\cite{Stephanibook, Griffiths_Podolský_2009} for $n=4$). As follows from~\eqref{Li0_GKS}, the optical matrix and geodesicity of $\mathbf{k}$ are invariant under a GKS transformation. Therefore, these coordinates will be canonical also for the background vacuum metric $\mathbf{\bar g}$, which can thus be written as \cite{Podolsky:2008ec} 
\begin{align}
    \mathbf{\bar g}= g_{\a\beta}dx^\a dx^\beta + 2(e_\a+f_\a r)dx^\a du -2du dr + (\bar ar^2+\bar br+\bar c)du^2,
		\label{back_Kundt}
\end{align}
where $g_{\a\beta}$, $e_\a$, $f_\a$, $\bar a$, $\bar b$, and $\bar c$ do not depend on $r$. The function $\bar a$ and the base space metric $g_{\a\beta}$ are constrained by \cite{Podolsky:2008ec}
\be
	\bar a=\frac{1}{2}(f^\a f_\a+f^\a_{\phantom{\a}||\a})+\frac{2\Lambda}{n-2} , \qquad {}^s\!R_{\a\beta}=\frac{2\Lambda}{n-2} g_{\a\beta}+\frac{1}{2}f_\a f_\beta+f_{(\a||\beta)} ,
	\label{back_eqs}
\ee
(also constraining $f_\a$ \cite{OrtPra16}) with the double bar denoting covariant differentiation in the base space geometry, ${}^s\!R_{\a\beta}$ the Ricci tensor associated to the latter, and spatial indices are raised with the inverse of the spatial metric $g^{\a\beta}$. Additionally, the functions $\bar b$ and $\bar c$ must obey the Einstein equations~(73) and (76) of \cite{Podolsky:2008ec} (after setting the Maxwell field to be zero there), which for brevity we do not reproduce here.\footnote{Let us note that some of the Einstein equations presented in \cite{Podolsky:2008ec} are redundant \cite{Krtousetal12,OrtPra16,Ortaggio:2024say}.
Additionally, the intermediate eq.~(25,\cite{Podolsky:2008ec}), from which (76,\cite{Podolsky:2008ec}) is derived, misses a term $-\frac{1}{2}g^{rr}g_{u\a,rr}$ (see~\cite{Podolsky:2013qwa}). However, the error does not propagate to (76,\cite{Podolsky:2008ec}) since the missing term turns out to be ultimately zero.}

In these coordinates, the field~\eqref{F_Kundt_gen_frame} with~\eqref{alpha_Kundt_case} and \eqref{alpha_null} reads
\begin{align}
    \mathbf{F}= -\beta_{0,\a} dx^\a\wedge du ,
		\label{F_Kundt}
\end{align}
so that the only non-zero component of the energy-momentum tensor is $T_{uu}=g^{\a\beta}\beta_{0,\a}\beta_{0,\beta}$. Except for the $(uu)$ components, all the remaining Einstein equations thus take the same form both for the vacuum background $\mathbf{\bar g}$ and for the full metric $\mathbf{g}$ \cite{Podolsky:2008ec}.

The charging through the GKS transformation is now simply captured by the following split
\begin{align}
    g_{uu}=\bar g_{uu}- 2H ,
		\label{GKS_transf_Kundt}
\end{align}
with~\eqref{r_depend_H_Kundt}, resulting in the full electrovac metric
\begin{align}
    \mathbf{g}=\mathbf{\bar g}+(b_Hr+c_H)du^2 ,
		\label{general_Kundt_metric}
\end{align}
coupled to the null Maxwell field~\eqref{F_Kundt}. 

The $r$-independent functions $\beta_0$, $b_H$ and $c_H$ are constrained by the Einstein-Maxwell equations \cite{Podolsky:2008ec} as follows
\begin{align}
& \T\beta_0=0 , \\
& b_H= b_H(u), \label{b+H+eqn}\\
& -\T c_{H}+f^\a c_{H,\a}+ f^\a_{\phantom{\a}||\a}c_{H}-b_{H}[e^\a_{\phantom{\a}||\a}-(\ln \sqrt{g})_{,u}]= 2\kappa g^{\a\gamma}\beta_{0,\a}\beta_{0,\gamma} , \label{c_H_eqn}
\end{align}
where $\T$ denote the Laplace operator in the base space and $g$ the determinant of the spatial metric.

To summarize, the complete family of GKS-Kundt solutions charged by an aligned $\mathbf{A}$ is given by~\eqref{F_Kundt}, \eqref{general_Kundt_metric}--\eqref{c_H_eqn}, where the  background (vacuum) Kundt geometry is described by~\eqref{back_Kundt} with \eqref{back_eqs} (and the two remaining Einstein equations for $\bar b$ and $\bar c$ mentioned above, see \cite{Podolsky:2008ec}). Related comments can be found in Section~3 of \cite{OrtPra16}.

As an explicit example, let us mention electrovac waves of Weyl type~II propagating in (anti-)Nariai product spaces in four \cite{PodOrt03} and higher dimensions \cite{Krtousetal12} (this is an example in which the vacuum background $\mathbf{\Bar{g}_0}$ is not itself GKS, cf. Remark~\ref{rem_OC}). Similarly, one can charge the full class of Kundt vacua of type~III (obtained for $n=4$ with an arbitrary $\Lambda$ in \cite{GriDocPod04}, and for any $n$ with $\Lambda=0$ in \cite{Coleyetal06}, cf. also \cite{Coleyetal07,OrtPra16}), or the one of the types~N and O (constructed in four dimensions with $\Lambda$ \cite{OzsRobRoz85}, and in higher dimensions without \cite{Coleyetal06} and with $\Lambda$ \cite{Ortaggio:2024say}) -- however, the latter solutions reduce to KS spacetimes \cite{OrtPraPra09,Malek:2010mh,Ortaggio18cqg,Ortaggio:2024say}.

\subsection{Case $\bar\Lambda\neq\Lambda$: non-null fields}

\label{subsec_app_Kundt_nonnull}

In this concluding section, we relax the assumption $\bar\Lambda=\Lambda$, which has been maintained throughout the rest of the paper. As we now show, this allows for GKS-Kundt solutions with a non-null $\mathbf{F}$. We still assume~\eqref{GKS}, \eqref{aligned_A}, with $\mathbf{k}$ Kundt w.r.t. $\mathbf{\bar g}$ (and thus, automatically, also w.r.t. $\mathbf{g}$, cf. \eqref{Li0_GKS}). 
The vacuum background is again described by~\eqref{back_Kundt} \cite{Podolsky:2008ec}, but now with \eqref{back_eqs} obviously replaced by
\be
	\bar a=\frac{1}{2}(f^\a f_\a +f^\a _{\phantom{\a }||\a })+\frac{2\bar\Lambda}{n-2} , \qquad {}^s\!R_{\a\beta}=\frac{2\bar\Lambda}{n-2} g_{\a\beta}+\frac{1}{2}f_\a f_\beta+f_{(\a ||\beta)} ,
	\label{back_eqs2}
\ee
while the Maxwell field is still given by~\eqref{F_Kundt_gen_frame} with~\eqref{F_compts} and \eqref{alpha_Kundt_case}.

For the full spacetimes, the components $(00)$ and $(0i)$ of the Einstein equation are identically satisfied (cf. eqs.~\eqref{R00_GKS}, \eqref{R0i_GKS}). 
The Einstein equation~\eqref{ij_eqn}, \eqref{01_eqn} with \eqref{Rij_GKS}, \eqref{R01_GKS}, \eqref{alpha_Kundt_case} then result in
\beqn
   & & {\kappa}\alpha_0^2=2(\bar\Lambda-\Lambda) , \label{alpha0_eqn_Kundt} \\
	 & & H= -\frac{1}{2}(-\kappa \alpha_0^2 r^2 + b_H r+ c_H) ,
\eeqn
where $b_H$ and $c_H$ do not depend on $r$. We thus see that $\alpha_0$ is a non-zero constant, since here we have assumed $\bar\Lambda\neq\Lambda$. Therefore $\mathbf{F}$ is necessarily non-null, as opposed to Section~\ref{subsec_app_Kundt_null}, and can be written as
\begin{align}
    \mathbf{F}
= -(\alpha_0 dr + \beta_{0,\a} dx^\a) \wedge du .
\label{F_Kundt_gen}
\end{align}

For the rest of the analysis, let us directly use the remaining field equations derived in \cite{Podolsky:2008ec}, namely, the Maxwell equations (79) and (80) of \cite{Podolsky:2008ec} and the Einstein equations (73) and (76) of \cite{Podolsky:2008ec}. Eq.~(80,\cite{Podolsky:2008ec}) simplifies in our case to $\alpha_0 f_\a=0$, consequently
\be
 f_\a=0 ,
 \label{fi=0}
\ee
which restricts the background space such that $\mathbf{k}$ is {\em recurrent} \cite{Stephanibook}, while eq.~(79,\cite{Podolsky:2008ec}) takes the form
\be
	\alpha_0 (\ln\sqrt{g})_{,u}= (\alpha_0 e^\gamma+\beta_0^{\phantom{0},\gamma})_{||\gamma} . 
	\label{Max_Kundt}
\ee	

Eq.~(76,\cite{Podolsky:2008ec}) reduces to
\be
 b_{H,\gamma}=-2\kappa\alpha_0\beta_{0,\gamma} ,
 \label{einst_bH_2}
\ee
such that $b_{H}=-2\kappa\alpha_0\beta_{0}+s_0(u)$, while (73,\cite{Podolsky:2008ec}) can be written as (using also~\eqref{Max_Kundt} and \eqref{einst_bH_2})
\be
	-\T c_{H}+\alpha_0^{-1}b_H{\T \beta_{0}} = 2\kappa g^{\a\gamma}\beta_{0,\a}\beta_{0,\gamma} . 
	\label{einst_cH_2}
\ee

Thus, the full family of GKS-Kundt metrics with Kundt $\mathbf{k}$ and aligned $\mathbf{A}$ with a non-null electromagnetic field~\eqref{F_Kundt_gen} (where $\alpha_0\neq0$) is given by
\begin{align}
    \mathbf{g}=\mathbf{\bar g}+(-\kappa \alpha_0^2 r^2 + b_H r+ c_H)du^2 ,
		\label{general_Kundt_metric2}
\end{align}
subject to the field equations~\eqref{alpha0_eqn_Kundt}, \eqref{Max_Kundt}, \eqref{einst_bH_2} and \eqref{einst_cH_2}. The background (vacuum) Kundt geometry $\mathbf{\bar g}$ is described by~\eqref{back_Kundt} with \eqref{back_eqs2}, \eqref{fi=0}, i.e., the transverse spatial metric defines an Einstein space with a cosmological constant $(n-4)\bar\Lambda/(n-2)$ (see \cite{Podolsky:2008ec} for the two remaining Einstein equations for $\bar b$ and $\bar c$, and for possible coordinate freedoms).

For example, a GKS-transformation of the above type can be used to produce the well-known charged direct product metrics \cite{LeviCivita17BR,Bertotti59,Robinson59,CahDef68} starting from an appropriate vacuum direct product metric \cite{Kasner25,Nariai50,CahDef68} (which include Minkowski spacetime) --  cf., e.g., eq.~(36,\cite{PodOrt03}) --
 also in the presence of gravitational waves \cite{Khlebnikov86,GarAlvar84,Lewand92,Ortaggio02,OrtPod02,PodOrt03} and in higher dimensions \cite{FreRub80,CarDiaLem04,Krtousetal12}.

%
%
%

\begin{thebibliography}{100}

\bibitem{Kerr63}
R.~P. Kerr, ``Gravitational field of a spinning mass as an example of
  algebraically special metrics,'' {\em Phys. Rev. Lett.}, vol.~11,
  pp.~237--238, 1963.

\bibitem{Newmanetal65}
E.~T. Newman, R.~Couch, K.~Chinnapared, A.~Exton, A.~Prakash, and R.~Torrence,
  ``Metric of a rotating, charged mass,'' {\em J. Math. Phys.}, vol.~6,
  pp.~918--919, 1965.

\bibitem{Trautman62}
A.~Trautman, ``On the propagation of information by waves,'' in {\em Recent
  Developments in General Relativity}, pp.~459--463, New York and Warszawa:
  Pergamon Press and PWN, 1962.

\bibitem{KerSch652}
R.~P. Kerr and A.~Schild, ``Some algebraically degenerate solutions of
  {E}instein's gravitational field equations,'' {\em Proc. Symp. Appl. Math.},
  vol.~17, pp.~199--209, 1965.

\bibitem{DebKerSch69}
G.~C. Debney, R.~P. Kerr, and A.~Schild, ``Solutions of the {E}instein and
  {E}instein-{M}axwell equations,'' {\em J. Math. Phys.}, vol.~10,
  pp.~1842--1854, 1969.

\bibitem{Stephanibook}
H.~Stephani, D.~Kramer, M.~MacCallum, C.~Hoenselaers, and E.~Herlt, {\em Exact
  Solutions of {E}instein's Field Equations}.
\newblock Cambridge: Cambridge University Press, second~ed., 2003.

\bibitem{Ayon-Beatoetal1025}
E.~Ay{\'o}n-Beato, D.~Flores-Alfonso, M.~Hassaine, and D.~F. Higuita-Borja,
  ``Dyonic {K}err-{S}child ansatz,'' {\em Phys. Rev. {\rm D}}, vol.~112,
  p.~104020, 2025.

\bibitem{Carter68cmp}
B.~Carter, ``{H}amilton-{J}acobi and {S}chrodinger separable solutions of
  {E}instein's equations,'' {\em Commun. Math. Phys.}, vol.~10, pp.~280--310,
  1968.

\bibitem{Carter73}
B.~Carter, ``Black hole equilibrium states,'' in {\em Black holes} (C.~De~Witt
  and B.~S. De~Witt, eds.), pp.~57--214, New York: Gordon and Breach, 1973.

\bibitem{Ort_Srini}
M.~Ortaggio and A.~Srinivasan, ``{Charging Kerr-Schild spacetimes in higher
  dimensions},'' {\em Phys. Rev. D}, vol.~110, no.~4, p.~044035, 2024.

\bibitem{Thompson66}
A.~H. Thompson, ``A class of related space-times,'' {\em Tensor}, vol.~17,
  pp.~92--95, 1966.

\bibitem{Edelen66_I}
D.~G.~B. Edelen, ``The null-bundle of an {E}instein-{R}iemann space~{I}:
  General theory,'' {\em J. Math. Mech.}, vol.~16, pp.~351--363, 1966.

\bibitem{Dozmorov70}
I.~M. Dozmorov, ``Solutions of the {E}instein equations with zero coupling,''
  {\em Soviet Physics Journal}, vol.~13, pp.~1284--1288, 1970.

\bibitem{GurGur75}
M.~G{\"{u}}rses and F.~G{\"{u}}rsey, ``{L}orentz covariant treatment of the
  {K}err-{S}child geometry,'' {\em J. Math. Phys.}, vol.~16, pp.~2385--2390,
  1975.

\bibitem{Xanthopoulos78}
B.~C. Xanthopoulos, ``Exact vacuum solutions of {E}instein's equation from
  linearized solutions,'' {\em J. Math. Phys.}, vol.~19, pp.~1607--1609, 1978.

\bibitem{Taub81}
A.~H. Taub, ``Generalized {K}err--{S}child space-times,'' {\em Ann. Phys.},
  vol.~134, pp.~326--372, 1981.

\bibitem{bilge}
A.~H. Bilge and M.~G{\"u}rses, ``Generalized {K}err-{S}child transformation,''
  in {\em Group Theoretical Methods in Physics} (M.~Serdaro{\u{g}}lu and
  E.~{\'I}n{\"o}n{\"u}, eds.), (Berlin, Heidelberg), pp.~252--255, Springer
  Berlin Heidelberg, 1983.

\bibitem{Xanthopoulos83}
B.~C. Xanthopoulos, ``The optical scalars in {K}err--{S}child-type
  spacetimes,'' {\em Ann. Physics}, vol.~149, pp.~286--295, 1983.

\bibitem{MarSen86}
J.~Mart\'{\i}n and J.~M.~M. Senovilla, ``Petrov type {D} perfect-fluid
  solutions in generalized {K}err-{S}child form,'' {\em J. Math. Phys.},
  vol.~27, pp.~265--270, 1986.

\bibitem{BilGur86}
A.~H. Bilge and M.~G{\"u}rses, ``Cartan ideal, prolongation, and {B}{\"a}cklund
  transformations for {E}instein's equations,'' {\em J. Math. Phys.},
  vol.~27, pp.~1819--1833, 1986.

\bibitem{Nahmad88}
E.~Nahmad-Achar, ``On generalized kerr-schild transformations,'' {\em J. Math.
  Phys.}, vol.~29, pp.~1879--1884, 1988.

\bibitem{ColHilSen01}
B.~Coll, S.~R. Hildebrandt, and J.~M.~M. Senovilla, ``Kerr-{S}child
  symmetries,'' {\em Gen. Rel. Grav.}, vol.~33, pp.~649--670, 2001.

\bibitem{MyePer86}
R.~C. Myers and M.~J. Perry, ``Black holes in higher dimensional space-times,''
  {\em Ann. Phys. (N.Y.)}, vol.~172, pp.~304--347, 1986.

\bibitem{Chakrabarti86}
A.~Chakrabarti, ``{K}err metric in eight dimensions,'' {\em Phys. Lett. {\rm
  B}}, vol.~172, pp.~175--179, 1986.

\bibitem{HawHunTay99}
S.~W. Hawking, C.~J. Hunter, and M.~M. Taylor-Robinson, ``Rotation and the
  {AdS/CFT} correspondence,'' {\em Phys. Rev. {\rm D}}, vol.~59, p.~064005,
  1999.

\bibitem{Gibbonsetal05}
G.~W. Gibbons, H.~L{\"u}, D.~N. Page, and C.~N. Pope, ``The general
  {K}err-de~{S}itter metrics in all dimensions,'' {\em J. Geom. Phys.},
  vol.~53, pp.~49--73, 2005.

\bibitem{MarPeo22_b}
M.~Mars and C.~Pe\'on-Nieto, ``Classification of {K}err--de~{S}itter-like
  spacetimes with conformally flat $\mathcal{I}$ in all dimensions,'' {\em
  Phys. Rev. {\rm D}}, vol.~105, p.~044027, 2022.

\bibitem{ChrConGra25}
P.~T. Chru\'{s}ciel, W.~Cong, and F.~Gray, ``{K}err-{A}d{S} type higher
  dimensional black holes with non-spherical cross-sections of horizons,'' {\em
  Class. Quantum Grav.}, vol.~42, p.~155007, 2025.

\bibitem{KleMorVan98}
D.~Klemm, V.~Moretti, and L.~Vanzo, ``Rotating topological black holes,'' {\em
  Phys. Rev. {\rm D}}, vol.~57, pp.~6127--6137, 1998.
\newblock See also D. Klemm, V. Moretti, and L. Vanzo (1999), Erratum: Rotating
  topological black holes [Phys. Rev. D 57, 6127 (1998)], {\em Phys. Rev.} D
  60:109902.

\bibitem{Klemm98}
D.~Klemm, ``Rotating black branes wrapped on {E}instein spaces,'' {\em JHEP},
  vol.~11, p.~019, 1998.

\bibitem{deFGodRea15}
G.~B. de~Freitas, M.~Godazgar, and H.~S. Reall, ``Uniqueness of the
  {K}err-de~{S}itter spacetime as an algebraically special solution in five
  dimensions,'' {\em Commun. Math. Phys.}, vol.~340, pp.~291--323, 2015.

\bibitem{Ortaggio17}
M.~Ortaggio, ``On the uniqueness of the {M}yers-{P}erry spacetime as a type
  {II(D)} solution in six dimensions,'' {\em JHEP}, vol.~06, p.~042, 2017.

\bibitem{MarPaeSen17}
M.~Mars, T.-T. Paetz, and J.~Senovilla, ``Classification of
  {K}err--de~{S}itter-like spacetimes with conformally flat $\mathcal{I}$,''
  {\em Class. Quantum Grav.}, vol.~34, p.~095010, 2017.

\bibitem{KokOrt25}
D.~Koko\v{s}ka and M.~Ortaggio, ``On the uniqueness of the {K}err-{(A)dS}
  metric as a type {II(D)} solution in six dimensions,'' {\em Phys. Rev. {\rm
  D}}, vol.~112, p.~044050, 2025.

\bibitem{DerGur86}
T.~Dereli and M.~G{\"{u}}rses, ``The generalized {K}err-{S}child transform in
  eleven-dimensional supergravity,'' {\em Phys. Lett. {\rm B}}, vol.~171,
  pp.~209--211, 1986.

\bibitem{OrtPraPra09}
M.~Ortaggio, V.~Pravda, and A.~Pravdov\'a, ``Higher dimensional {K}err-{S}child
  spacetimes,'' {\em Class. Quantum Grav.}, vol.~26, p.~025008, 2009.

\bibitem{Malek:2010mh}
T.~Malek and V.~Pravda, ``{Kerr-Schild spacetimes with (A)dS background},''
  {\em Class. Quant. Grav.}, vol.~28, p.~125011, 2011.

\bibitem{Srinivasan:2025hro}
A.~Srinivasan, ``{Algebraic and optical properties of generalized Kerr-Schild
  spacetimes in arbitrary dimensions},'' {\em Phys. Rev. D}, vol.~111, no.~6,
  p.~064061, 2025.

\bibitem{BerSen01}
G.~Bergqvist and J.~Senovilla, ``Null cone preserving maps, causal tensors and
  algebraic {R}ainich theory,'' {\em Class. Quantum Grav.}, vol.~18,
  pp.~5299--5326, 2001.

\bibitem{Coleyetal04vsi}
A.~Coley, R.~Milson, V.~Pravda, and A.~Pravdov\'a, ``Vanishing scalar invariant
  spacetimes in higher dimensions,'' {\em Class. Quantum Grav.}, vol.~21,
  pp.~5519--5542, 2004.

\bibitem{Milson05}
R.~Milson, ``Alignment and the classification of {L}orentz-signature tensors,''
  in {\em Symmetry and Perturbation Theory: Proceedings of the International
  Conference on SPT 2004} (G.~Gaeta, B.~Prinari, S.~Rauch-Wojciechowski, and
  S.~Terracini, eds.), (Singapore), pp.~215--222, World Scientific, 2005. [gr-qc/0411036]

\bibitem{HerOrtWyl13}
S.~Hervik, M.~Ortaggio, and L.~Wylleman, ``Minimal tensors and purely electric
  or magnetic spacetimes of arbitrary dimension,'' {\em Class. Quantum Grav.},
  vol.~30, p.~165014, 2013.

\bibitem{Harte17}
A.~I. Harte, ``Metric-independence of vacuum and force-free electromagnetic
  fields,'' {\em Phys. Rev. Lett.}, vol.~118, p.~141101, 2017.

\bibitem{Kupeli88}
A.~H. Kupeli, ``Generalised {K}err-{S}child transformation from vacuum
  backgrounds to {E}instein-{M}axwell spacetimes,'' {\em Class. Quantum Grav.},
  vol.~5, pp.~401--408, 1988.

\bibitem{Awad03}
A.~M. Awad, ``Higher dimensional charged rotating solutions in {(A)dS}
  space-times,'' {\em Class. Quantum Grav.}, vol.~20, pp.~2827--2833, 2003.

\bibitem{Lemos95}
J.~P.~S. Lemos, ``Two-dimensional black holes and planar general relativity,''
  {\em Class. Quantum Grav.}, vol.~12, pp.~1081--1086, 1995.

\bibitem{LemZan96}
J.~P.~S. Lemos and V.~T. Zanchin, ``Rotating charged black string and
  three-dimensional black holes,'' {\em Phys. Rev. {\rm D}}, vol.~54,
  pp.~3840--3853, 1996.

\bibitem{GibWil87}
G.~W. Gibbons and D.~L. Wiltshire, ``Space-time as a membrane in higher
  dimensions,'' {\em Nucl. Phys. {\rm B}}, vol.~287, pp.~717--742, 1987.

\bibitem{KodIsh04}
H.~Kodama and A.~Ishibashi, ``Master equations for perturbations of generalized
  static black holes with charge in higher dimensions,'' {\em Prog. Theor.
  Phys.}, vol.~111, pp.~29--73, 2004.

\bibitem{OrtPodZof08}
M.~Ortaggio, J.~Podolsk\'y, and M.~\v{Z}ofka, ``{R}obinson-{T}rautman
  spacetimes with an electromagnetic field in higher dimensions,'' {\em Class.
  Quantum Grav.}, vol.~25, p.~025006, 2008.

\bibitem{KunNavPer05}
J.~Kunz, F.~Navarro-L\'erida, and A.~K. Petersen, ``Five-dimensional charged
  rotating black holes,'' {\em Phys. Lett. {\rm B}}, vol.~614, pp.~104--112,
  2005.

\bibitem{KunNavVie06}
J.~Kunz, F.~Navarro-L\'erida, and J.~Viebahn, ``Charged rotating black holes in
  odd dimensions,'' {\em Phys. Lett. {\rm B}}, vol.~639, pp.~362--367, 2006.

\bibitem{Frobetal22}
M.~B. Fr{\"o}b, I.~Khavkine, T.~M{\'a}lek, and V.~Pravda, ``On well-posedness
  and algebraic type of the five-dimensional charged rotating black hole with
  two equal-magnitude angular momenta,'' {\em Eur. Phys. J.~C}, vol.~82,
  p.~215, 2022.

\bibitem{Breckenridgeetal97}
J.~C. Breckenridge, R.~C. Myers, A.~W. Peet, and C.~Vafa, ``{D}-branes and
  spinning black holes,'' {\em Phys. Lett. {\rm B}}, vol.~391, pp.~93--98,
  1997.

\bibitem{CCLP}
Z.-W. Chong, M.~Cveti\ifmmode~\check{c}\else \v{c}\fi{}, H.~L\"u, and C.~N.
  Pope, ``General nonextremal rotating black holes in minimal five-dimensional
  gauged supergravity,'' {\em Phys. Rev. Lett.}, vol.~95, p.~161301, Oct 2005.

\bibitem{Elvangetal04}
H.~Elvang, R.~Emparan, D.~Mateos, and H.~S. Reall, ``A supersymmetric black
  ring,'' {\em Phys. Rev. Lett.}, vol.~93, p.~211302, 2004.

\bibitem{DesLun25}
R.~Deshpande and O.~Lunin, ``Rotating {E}instein-{M}axwell black holes in odd
  dimensions,'' {\em JHEP}, vol.~06, p.~066, 2025.

\bibitem{EmpRea08}
R.~Emparan and H.~S. Reall, ``Black holes in higher dimensions,'' {\em Living
  Reviews in Relativity}, vol.~11, no.~6, 2008.

\bibitem{Taghavi-Chabert22}
A.~Taghavi-Chabert, ``Twisting non-shearing congruences of null geodesics,
  almost {CR} structures and {E}instein metrics in even dimensions,'' {\em Ann.
  Mat. Pura Appl.}, vol.~201, pp.~655--693, 2022.

\bibitem{ManSte06}
R.~Mann and C.~Stelea, ``New {T}aub-{NUT}-{R}eissner-{N}ordstr{\"o}m spaces in
  higher dimensions,'' {\em Phys. Lett. {\rm B}}, vol.~632, pp.~537--542, 2006.

\bibitem{Awad06}
A.~M. Awad, ``Higher dimensional {T}aub-nuts and {T}aub-bolts in
  {E}instein-{M}axwell gravity,'' {\em Class. Quantum Grav.}, vol.~23,
  pp.~2849--2859, 2006.

\bibitem{DehKoh06}
M.~H. Dehghani and A.~Khodam-Mohammadi, ``Thermodynamics of {T}aub-{NUT}/bolt
  black holes in {E}instein-{M}axwell gravity,'' {\em Phys. Rev. {\rm D}},
  vol.~73, p.~124039, 2006.

\bibitem{type_III_N}
M.~Ortaggio, V.~Pravda, and A.~Pravdova, ``{Type III and N Einstein spacetimes
  in higher dimensions: General properties},'' {\em Phys. Rev. D}, vol.~82,
  p.~064043, 2010.

\bibitem{OrtPraPra13}
M.~Ortaggio, V.~Pravda, and A.~Pravdov\'a, ``On the {G}oldberg-{S}achs theorem
  in higher dimensions in the non-twisting case,'' {\em Class. Quantum Grav.},
  vol.~30, p.~075016, 2013.

\bibitem{OrtPraPra13rev}
M.~Ortaggio, V.~Pravda, and A.~Pravdov\'a, ``Algebraic classification of higher
  dimensional spacetimes based on null alignment,'' {\em Class. Quantum Grav.},
  vol.~30, p.~013001, 2013.

\bibitem{Coleyetal04}
A.~Coley, R.~Milson, V.~Pravda, and A.~Pravdov\'a, ``Classification of the
  {W}eyl tensor in higher dimensions,'' {\em Class. Quantum Grav.}, vol.~21,
  pp.~L35--L41, 2004.

\bibitem{Pravdaetal04}
V.~Pravda, A.~Pravdov\'a, A.~Coley, and R.~Milson, ``Bianchi identities in
  higher dimensions,'' {\em Class. Quantum Grav.}, vol.~21, pp.~2873--2897,
  2004.
\newblock See also V. Pravda, A. Pravdov\'a, A. Coley and R. Milson {\em Class.
  Quantum Grav.} {\bf 24} (2007) 1691 (corrigendum).

\bibitem{OrtPraPra07}
M.~Ortaggio, V.~Pravda, and A.~Pravdov\'a, ``Ricci identities in higher
  dimensions,'' {\em Class. Quantum Grav.}, vol.~24, pp.~1657--1664, 2007.

\bibitem{DurRea09}
M.~Durkee and H.~S. Reall, ``A higher-dimensional generalization of the
  geodesic part of the {G}oldberg-{S}achs theorem,'' {\em Class. Quantum
  Grav.}, vol.~26, p.~245005, 2009.

\bibitem{OrtPraPra09b}
M.~Ortaggio, V.~Pravda, and A.~Pravdov\'a, ``Asymptotically flat, algebraically
  special spacetimes in higher dimensions,'' {\em Phys. Rev. {\rm D}}, vol.~80,
  p.~084041, 2009.

\bibitem{Ortaggioetal12}
M.~Ortaggio, V.~Pravda, A.~Pravdov\'a, and H.~S. Reall, ``On a five-dimensional
  version of the {G}oldberg-{S}achs theorem,'' {\em Class. Quantum Grav.},
  vol.~29, p.~205002, 2012.

\bibitem{OrtPraPra18}
M.~Ortaggio, V.~Pravda, and A.~Pravdov\'a, ``On higher dimensional {E}instein
  spacetimes with a non-degenerate double {W}eyl aligned null direction,'' {\em
  Class. Quantum Grav.}, vol.~35, p.~075004, 2018.

\bibitem{TinPra19}
T.~Tint\v{e}ra and V.~Pravda, ``On the {G}oldberg--{S}achs theorem in six
  dimensions,'' {\em Gen. Rel. Grav.}, vol.~51, p.~111, 2019.

\bibitem{Tangherlini63}
F.~R. Tangherlini, ``Schwarzschild field in $n$ dimensions and the
  dimensionality of space problem,'' {\em Il Nuovo Cimento}, vol.~27,
  pp.~636--651, 1963.

\bibitem{Podolsky:2008ec}
J.~Podolsk\'y and M.~\v{Z}ofka, ``{General Kundt spacetimes in higher
  dimensions},'' {\em Class. Quant. Grav.}, vol.~26, p.~105008, 2009.

\bibitem{Berard82}
L.~B\'erard~Bergery, ``Sur de nouvelles vari{\'e}t{\'e}s riemanniennes
  d'{E}instein,'' in {\em Institut {E}lie {C}artan}, vol.~6, pp.~1--60, Nancy:
  Equipe de recherche associ{\'e}e au {CNRS} d'Analyse Globale n$^o$~839,
  {U}niversit{\'e} de Nancy~{I}, 1982.

\bibitem{Bais:1984xb}
F.~A. Bais and P.~Batenburg, ``{A New Class of Higher Dimensional
  {Kaluza-Klein} Monopole and Instanton Solutions},'' {\em Nucl. Phys. B},
  vol.~253, pp.~162--172, 1985.

\bibitem{PagPop87}
D.~N. Page and C.~N. Pope, ``Inhomogeneous {E}instein metrics on complex line
  bundles,'' {\em Class. Quantum Grav.}, vol.~4, pp.~213--225, 1987.

\bibitem{Lor_Dieter}
D.~Lorenz-Petzold, ``{Higher-Dimensional Taub-NUT-De Sitter Solutions},'' {\em
  Progress of Theoretical Physics}, vol.~78, pp.~11--15, 07 1987.

\bibitem{ChaGib96}
A.~Chamblin and G.~W. Gibbons, ``Topology and time reversal,'' in {\em String
  Gravity and Physics at the {P}lanck Energy Scale} (N.~S\'anchez and
  A.~Zichichi, eds.), vol.~476, pp.~233--253, Dordrecht/Boston/London: Kluwer
  Academic Publishers, 1996.

\bibitem{Taylor:1998fd}
M.~Taylor, ``{Higher dimensional Taub-Bolt solutions and the entropy of
  noncompact manifolds},'' hep-th/9809041, 1998.

\bibitem{Awad:2000gg}
A.~Awad and A.~Chamblin, ``{A Bestiary of higher dimensional Taub - NUT AdS
  space-times},'' {\em Class. Quant. Grav.}, vol.~19, pp.~2051--2062, 2002.

\bibitem{Mann:Nuttier}
R.~B. Mann and C.~Stelea, ``{Nuttier (A)dS black holes in higher dimensions},''
  {\em Class. Quant. Grav.}, vol.~21, pp.~2937--2962, 2004.

\bibitem{DesJacTem82}
S.~Deser, R.~Jackiw, and S.~Templeton, ``Topologically massive gauge
  theories,'' {\em Ann. Phys.}, vol.~140, pp.~372--411, 1982.
\newblock See also S. Deser and R. Jackiw (1988), Erratum, {\em Ann. Phys.}
  185:406.

\bibitem{GauMyeTow99}
J.~P. Gauntlett, R.~C. Myers, and P.~K. Townsend, ``Black holes of {D=5}
  supergravity,'' {\em Class. Quantum Grav.}, vol.~16, pp.~1--21, 1999.

\bibitem{Durkeeetal10}
M.~Durkee, V.~Pravda, A.~Pravdov\'a, and H.~S. Reall, ``Generalization of the
  {G}eroch-{H}eld-{P}enrose formalism to higher dimensions,'' {\em Class.
  Quantum Grav.}, vol.~27, p.~215010, 2010.

\bibitem{Ortaggio:2014ipa}
M.~Ortaggio, ``{Asymptotic behavior of Maxwell fields in higher dimensions},''
  {\em Phys. Rev. D}, vol.~90, no.~12, p.~124020, 2014.

\bibitem{Ortaggio09}
M.~Ortaggio, ``{B}el-{D}ebever criteria for the classification of the {W}eyl
  tensor in higher dimensions,'' {\em Class. Quantum Grav.}, vol.~26,
  p.~195015, 2009.

\bibitem{Hassaine:2024mfs}
M.~Hassaine, D.~Kubiz{\v n}{\'a}k, and A.~Srinivasan, ``{Extremal Kerr-Schild form},''
  {\em Phys. Rev. D}, vol.~111, no.~6, p.~L061502, 2025.

\bibitem{Malek_xKS}
T.~M\'alek, ``{Extended Kerr-Schild spacetimes: General properties and some
  explicit examples},'' {\em Class. Quant. Grav.}, vol.~31, p.~185013, 2014.

\bibitem{Carter68pla}
B.~Carter, ``A new family of {E}instein spaces,'' {\em Phys. Lett. {\rm A}},
  vol.~26, pp.~399--400, 1968.

\bibitem{Gold_Sachs_62}
J.~N. {Goldberg} and R.~K. {Sachs}, ``{A theorem on Petrov types},'' {\em Acta
  Physica Polonica B, Proceedings Supplement}, vol.~22, p.~13, Jan. 1962.

\bibitem{LuPagPop04}
H.~L{\"u}, D.~N. Page, and C.~N. Pope, ``New inhomogeneous {E}instein metrics
  on sphere bundles over {E}instein-{K}{\"a}hler manifolds,'' {\em Phys. Lett.
  {\rm B}}, vol.~593, pp.~218--226, 2004.

\bibitem{multi_NUT_Mann}
R.~B. Mann and C.~Stelea, ``{New multiply nutty spacetimes},'' {\em Phys. Lett.
  B}, vol.~634, pp.~448--455, 2006.

\bibitem{Chen_Pope_Kerr_NUT_multi}
W.~Chen, H.~Lu, and C.~N. Pope, ``{General Kerr-NUT-AdS metrics in all
  dimensions},'' {\em Class. Quant. Grav.}, vol.~23, pp.~5323--5340, 2006.

\bibitem{Pravda_type_D}
V.~Pravda, A.~Pravdova, and M.~Ortaggio, ``{Type D Einstein spacetimes in
  higher dimensions},'' {\em Class. Quant. Grav.}, vol.~24, pp.~4407--4428,
  2007.

\bibitem{Srinivasan_PhD}
A.~Srinivasan.
\newblock {PhD} thesis, Charles University, Prague.
\newblock To appear.

\bibitem{Gibbons_rot_HD_cosmol}
G.~W. Gibbons, H.~Lu, D.~N. Page, and C.~N. Pope, ``{Rotating black holes in
  higher dimensions with a cosmological constant},'' {\em Phys. Rev. Lett.},
  vol.~93, p.~171102, 2004.

\bibitem{FloQue19}
D.~Flores-Alfonso and H.~Quevedo, ``Topological characterization of
  higher-dimensional charged {T}aub-{NUT} instantons,'' {\em Int. J. Geom.
  Meth. Mod. Phys.}, vol.~16, p.~1950154, 2019.

\bibitem{Brill64}
D.~R. Brill, ``Electromagnetic fields in a homogeneous, nonisotropic
  universe,'' {\em Phys. Rev.}, vol.~133, pp.~845--848, 1964.

\bibitem{Ruban72}
V.~A. Ruban, ``Non-singular metrics of {T}aub-{N}ewman-{U}nti-{T}amburino type
  with an electromagnetic field,'' {\em Dokl. Akad. Nauk SSSR}, vol.~204,
  pp.~1086--1089, 1972.
\newblock In Russian.

\bibitem{Bochner1947CurvatureIH}
S.~Bochner, ``Curvature in hermitian metric,'' {\em Bulletin of the American
  Mathematical Society}, vol.~53, pp.~179--195, 1947.

\bibitem{yano1965differential}
K.~Yano, {\em Differential Geometry on Complex and Almost Complex Spaces}.
\newblock International series of monographs in pure and applied mathematics,
  Macmillan, 1965.

\bibitem{Okubo_1970}
S.~Kobayashi and K.~Nomizu, {\em Foundations of Differential Geometry, vol. 2}.
\newblock Interscience Publication, Wiley, New York, 1969.

\bibitem{1999CQGra..16L...9N}
P.~{Nurowski} and M.~{Przanowski}, ``{A four-dimensional example of a Ricci
  flat metric admitting almost-K{\"a}hler non-K{\"a}hler structure},'' {\em
  Classical and Quantum Gravity}, vol.~16, pp.~L9--L13, Mar. 1999.

\bibitem{Armstrong2002AnAF}
J.~Armstrong, ``An ansatz for almost-{K}{\"a}hler, {E}instein 4-manifolds,'' {\em
  J. reine angew. Math.}, vol.~2002, no. 542, 2002.

\bibitem{Robinson:1960zzb}
I.~Robinson and A.~Trautman, ``{Spherical Gravitational Waves},'' {\em Phys.
  Rev. Lett.}, vol.~4, pp.~431--432, 1960.

\bibitem{Robinson:1962zz}
I.~Robinson and A.~Trautman, ``{Some spherical gravitational waves in general
  relativity},'' {\em Proc. Roy. Soc. Lond. A}, vol.~265, pp.~463--473, 1962.

\bibitem{Griffiths_Podolský_2009}
J.~B. Griffiths and J.~Podolsk\'y, {\em Exact Space-Times in Einstein's
  General Relativity}.
\newblock Cambridge Monographs on Mathematical Physics, Cambridge University
  Press, 2009.

\bibitem{Podolsky_RT}
J.~Podolsk\'y and M.~Ortaggio, ``{Robinson-Trautman spacetimes in higher
  dimensions},'' {\em Class. Quant. Grav.}, vol.~23, pp.~5785--5797, 2006.

\bibitem{Reall:2012ih}
H.~S. Reall, A.~A.~H. Graham, and C.~P. Turner, ``{On algebraically special
  vacuum spacetimes in five dimensions},'' {\em Class. Quant. Grav.}, vol.~30,
  p.~055004, 2013.

\bibitem{Alekseevskyetal21}
D.~V. Alekseevsky, M.~Ganji, G.~Schmalz, and A.~Spiro, ``{L}orentzian manifolds
  with shearfree congruences and {K}{\"a}hler-{S}asaki geometry,'' {\em Differ.
  Geom. Appl.}, vol.~75, p.~101724, 2021.

\bibitem{Ortaggio:RT_pform}
M.~Ortaggio, J.~Podolsk\'y, and M.~\v{Z}ofka, ``{Static and radiating p-form
  black holes in the higher dimensional Robinson-Trautman class},'' {\em JHEP},
  vol.~02, p.~045, 2015.

\bibitem{ColHerPel06}
A.~Coley, S.~Hervik, and N.~Pelavas, ``On spacetimes with constant scalar
  invariants,'' {\em Class. Quantum Grav.}, vol.~23, pp.~3053--3074, 2006.

\bibitem{Coleyetal03}
A.~Coley, R.~Milson, N.~Pelavas, V.~Pravda, A.~Pravdov\'a, and R.~Zalaletdinov,
  ``Generalizations of \pp-wave spacetimes in higher dimensions,'' {\em Phys.
  Rev. {\rm D}}, vol.~67, p.~104020, 2003.

\bibitem{Podolsky:2013qwa}
J.~Podolsk\'y and R.~{\v{S}}varc, ``{Explicit algebraic classification of Kundt
  geometries in any dimension},'' {\em Class. Quant. Grav.}, vol.~30,
  p.~125007, 2013.

\bibitem{OrtPra16}
M.~Ortaggio and V.~Pravda, ``Electromagnetic fields with vanishing scalar
  invariants,'' {\em Class. Quantum Grav.}, vol.~33, p.~115010, 2016.

\bibitem{Krtousetal12}
P.~Krtou\v{s}, J.~Podolsk\'{y}, A.~Zelnikov, and H.~Kadlecov{\'a},
  ``Higher-dimensional {K}undt waves and gyratons,'' {\em Phys. Rev. {\rm D}},
  vol.~86, p.~044039, 2012.

\bibitem{Ortaggio:2024say}
M.~Ortaggio, J.~Vold{\v{r}}ich, and J.~Barrientos, ``{All nonexpanding
  gravitational waves in D-dimensional (anti{\textendash})de Sitter space},''
  {\em Phys. Rev. D}, vol.~110, no.~12, p.~124032, 2024.

\bibitem{PodOrt03}
J.~Podolsk\'y and M.~Ortaggio, ``Explicit {K}undt type {$II$} and {$N$}
  solutions as gravitational waves in various type {$D$} and {$O$} universes,''
  {\em Class. Quantum Grav.}, vol.~20, pp.~1685--1701, 2003.

\bibitem{GriDocPod04}
J.~Griffiths, P.~Docherty, and J.~Podolsk\'y, ``Generalized {K}undt waves and
  their physical interpretation,'' {\em Class. Quantum Grav.}, vol.~21,
  pp.~207--222, 2004.

\bibitem{Coleyetal06}
A.~Coley, A.~Fuster, S.~Hervik, and N.~Pelavas, ``Higher dimensional {VSI}
  spacetimes,'' {\em Class. Quantum Grav.}, vol.~23, pp.~7431--7444, 2006.

\bibitem{Coleyetal07}
A.~Coley, A.~Fuster, S.~Hervik, and N.~Pelavas, ``Vanishing scalar invariant
  spacetimes in supergravity,'' {\em JHEP}, vol.~0705, p.~032, 2007.

\bibitem{OzsRobRoz85}
I.~Ozsv\'ath, I.~Robinson, and K.~R\'ozga, ``Plane-fronted gravitational and
  electromagnetic waves in spaces with cosmological constant,'' {\em J. Math.
  Phys.}, vol.~26, pp.~1755--1761, 1985.

\bibitem{Ortaggio18cqg}
M.~Ortaggio, ``A note on {K}undt spacetimes of type {N} with a cosmological
  constant,'' {\em Class. Quantum Grav.}, vol.~35, p.~127001, 2018.

\bibitem{LeviCivita17BR}
T.~Levi-Civita, ``Realt\`a fisica di alcuni spazi normali del {B}ianchi,'' {\em
  Rend. Acc. Lincei}, vol.~26, pp.~519--531, 1917.

\bibitem{Bertotti59}
B.~Bertotti, ``Uniform electromagnetic field in the theory of general
  relativity,'' {\em Phys. Rev.}, vol.~116, pp.~1331--1333, 1959.

\bibitem{Robinson59}
I.~Robinson, ``A solution of the {M}axwell--{E}instein equations,'' {\em Bull.
  Acad. Polon.}, vol.~7, pp.~351--352, 1959.

\bibitem{CahDef68}
M.~Cahen and L.~Defrise, ``Lorentzian 4 dimensional manifolds with ``local
  isotropy'','' {\em Commun. Math. Phys.}, vol.~11, pp.~56--76, 1968.

\bibitem{Kasner25}
E.~Kasner, ``An algebraic solution of the {E}instein equations,'' {\em Trans.
  Am. Math. Soc.}, vol.~27, pp.~101--105, 1925.

\bibitem{Nariai50}
H.~Nariai, ``On some static solutions of {E}instein's gravitational field
  equations in a spherically symmetric case,'' {\em Sci. Rep. T\^{o}hoku
  Univ.}, vol.~34, pp.~160--167, 1950.

\bibitem{Khlebnikov86}
V.~I. Khlebnikov, ``Gravitational radiation in electromagnetic universes,''
  {\em Class. Quantum Grav.}, vol.~3, pp.~169--173, 1986.

\bibitem{GarAlvar84}
A.~Garc\'{\i}a~D. and M.~Alvarez~C., ``Shear-free special electrovac type-{II}
  solutions with cosmological constant,'' {\em Nuovo Cimento {\rm B}}, vol.~79,
  pp.~266--270, 1984.

\bibitem{Lewand92}
J.~Lewandowski, ``Reduced holonomy group and {E}instein equations with a
  cosmological constant,'' {\em Class. Quantum Grav.}, vol.~9, pp.~L147--L151,
  1992.

\bibitem{Ortaggio02}
M.~Ortaggio, ``Impulsive waves in the {N}ariai universe,'' {\em Phys. Rev. {\rm
  D}}, vol.~65, p.~084046, 2002.

\bibitem{OrtPod02}
M.~Ortaggio and J.~Podolsk\'y, ``Impulsive waves in electrovac direct product
  spacetimes with {$\Lambda$},'' {\em Class. Quantum Grav.}, vol.~19,
  pp.~5221--5227, 2002.

\bibitem{FreRub80}
P.~G.~O. Freund and M.~A. Rubin, ``Dynamics of dimensional reduction,'' {\em
  Phys. Lett. {\rm B}}, vol.~97, pp.~233--235, 1980.

\bibitem{CarDiaLem04}
V.~Cardoso, O.~J.~C. Dias, and J.~P.~S. Lemos, ``Nariai, {B}ertotti-{R}obinson
  and anti-{N}ariai solutions in higher dimensions,'' {\em Phys. Rev. {\rm D}},
  vol.~70, p.~024002, 2004.

\end{thebibliography}

\end{document}